\documentclass[reprint,superscriptaddress,amsmath,amssymb,aps,prl]{revtex4-1}
\usepackage{graphicx}
\usepackage{dcolumn}
\usepackage{bm}
\usepackage{color}
\usepackage{units}
\usepackage{comment}
\usepackage{siunitx}
\begin{document}

\title{Dynamic Overlap Concentration Scale of Active Colloids}

\author{Stewart A. Mallory}
\email{smallory@caltech.edu}
\affiliation{Division of Chemistry and Chemical Engineering, California Institute of Technology, Pasadena, California, 91125, USA}

\author{Ahmad K. Omar}
\email{aomar@berkeley.edu}
\affiliation{Department of Chemistry, University of California, Berkeley, California, 94720, USA }

\author{John F. Brady}
\email{jfbrady@caltech.edu}
\affiliation{Division of Chemistry and Chemical Engineering, California Institute of Technology, Pasadena, California, 91125, USA}

\begin{abstract}
By introducing the notion of a dynamic overlap concentration scale, we identify universal and previously unreported features of the mechanical properties of active colloids.
These features are codified by recognizing that the characteristic length scale of an active particle's trajectory, the run-length, introduces a new concentration scale $\phi^*$.
Large-scale simulations of repulsive active Brownian particles (ABPs) confirm that this new run-length dependent concentration, which is the trajectory-space analogue of the overlap concentration in polymer solutions, delineates distinct concentration regimes in which interparticle collisions alter particle trajectories.
Using $\phi^*$ and concentration scales associated with colloidal jamming, the mechanical equation-of-state for ABPs can be collapsed onto a set of principal curves that contain a number of previously overlooked features. 
The inclusion of these features qualitatively alters previous predictions of the behavior for active colloids as we demonstrate by computing the spinodal for a suspension of purely-repulsive ABPs.
Our findings suggest that dynamic overlap concentration scales should be of great utility in unraveling the behavior of active and driven systems.
\end{abstract}

\maketitle

\emph{Introduction.--} Pressure continues to be a topic of fundamental interest in active matter~\cite{Fily2014a, Mallory2014b, Takatori2014, Solon2015, solon_pressure_2015-1, Smallenburg2015, Winkler2015, speck_ideal_2016, Joyeux2016,fily2018, MariniBettoloMarconi2017, Patch2017, Das2019, Epstein2019, Omar2020}.
An interest motivated, in part, by the tantalizing prospect that mechanical equations-of-state (EoS) will play as important of a role in active matter as they do in equilibrium theory.
In fact, pressure has already proven to be a salient metric for rationalizing the often-complex behavior of active suspensions ranging from instabilities exhibited by expanding bacterial droplets~\cite{Sokolov2018}, vesicles filled with active particles~\cite{Li2019,Paoluzzi2016,Wang2019, Takatori2020}, active depletion~\cite{Harder2014,ZaeifiYamchi2017,Ni2015}, and the dynamics of colloidal gels~\cite{Mallory2020,Szakasits2017,Omar2019}, membranes~\cite{Mallory2015}, and polymers~\cite{Kaiser2014a,Kaiser2015,Xia2019} immersed in a bath of active colloids.
The burgeoning of research on the nature of pressure in active systems has not only contributed to progress in understanding the phenomenology of active systems, but has also played a central role in assessing the validity of new theoretical concepts for nonequilibrium systems~\cite{Takatori2016a,Lee2017,Rein2016,Zakine2020,marconi_pressure_2016,Speck2016,MariniBettoloMarconi2015,Wittmann2019,Marconi2017,Rodenburg2017,Chakraborti2016}.
There is no better example than the intense focus on the development of nonequilibrium theories for the phase behavior of active particles~\cite{Takatori2015,Paliwal2018,Wittkowski2014,Takatori2015a,Partridge2019,Solon2018,Hermann2019,Solon2015,Levis2017,Hermann2019a,Tjhung2018} ~{-- a crucial ingredient for many is an EoS}.

Pressure in active systems remains difficult to characterize analytically in all but the simplest cases.
These challenges can be attributed to a unique nonlocal contribution to the pressure referred to as the swim pressure~\cite{Fily2014a, Takatori2014}.
This nontraditional source of stress is entirely rooted in trajectory space and unraveling its concentration dependence necessitates the difficult task of understanding how particle trajectories are altered by many-body correlations.
The driving force for much of the novel behaviors observed in active systems, including the phenomena of motility-induced phase separation (MIPS), is due to this inherently nonequilibrium contribution to the pressure.

The swim pressure emerges from the persistent and time-irreversible motion generated by the non-conservative self-propelling force of active particles.
Within the context of the active Brownian particle (ABP) model – a popular minimal model for studying active colloids – particles self-propel at a constant speed $U_0$ and undergo Brownian rotational diffusion with a characteristic reorientation time $\tau_R$.
At long times, an ideal ABP executes a random walk with a trajectory correlation length $\ell_0=U_0 \tau_R$, which we call the intrinsic run-length. 
A dilute system of ABPs with number density $n$ exerts a swim pressure directly proportional to this intrinsic run-length, $\Pi^s \sim n \ell_0 U_0$.  
For more concentrated systems, interparticle interactions reduce the trajectory correlation length, resulting in a density-dependent effective run-length $\ell < \ell_0$ and a swim pressure of $\Pi^s \sim n \ell U_0$. 
In the limit of maximal-packing, ABPs are prevented from executing their athermal random walk resulting in a vanishingly small effective run-length (and swim pressure) that should be independent of $\ell_0$ and depend only on geometric packing constraints. 
This physical expectation for the swim pressure is currently not reflected in the literature. 

\begin{figure}[t!]
	\centering
	\includegraphics[width=.475\textwidth]{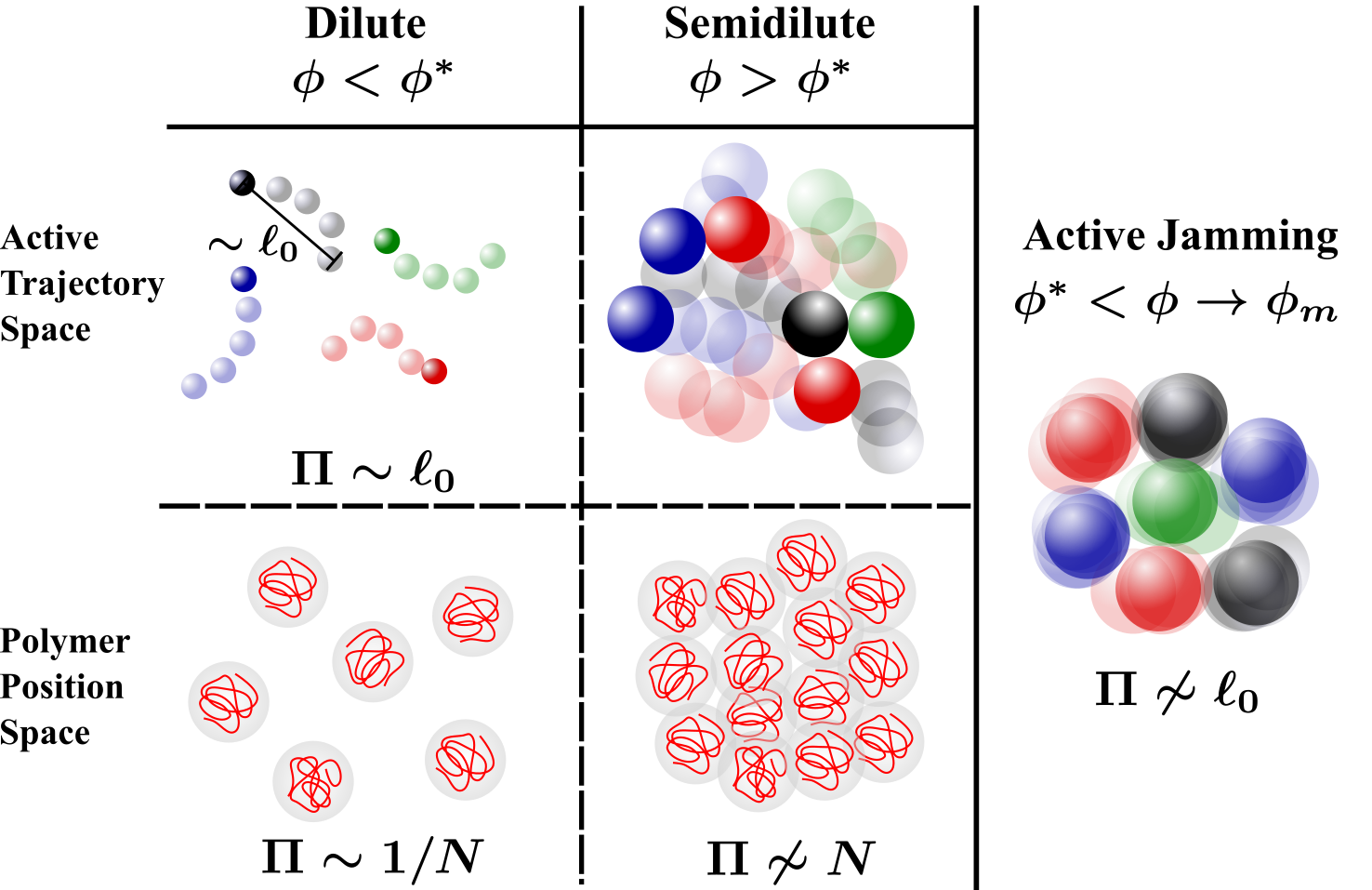}
	\caption{\protect\small{{The overlap concentration for polymer solutions and the proposed trajectory-space analogue for active colloids. Scaling  of the equilibrium osmotic pressure of the polymer solution with molecular weight $N$ and the pressure of active colloids with $\ell_0$ are shown in different concentration regimes. In the semidilute regime for both polymers and active colloids, the EoS can be collapsed onto a universal curve using an appropriately defined $\phi^*$. }}}
	\label{figure1}
\end{figure}

This simple physical picture [see Fig.~\ref{figure1}] suggests the existence of a concentration scale that controls the crossover between the dynamically distinct dilute and jammed regimes of active suspensions.
In this Letter, using large-scale computer simulations, we identify this new run-length dependent concentration $\phi^*$, which exhibits a number of compelling analogies to the powerful concept of the overlap concentration of equilibrium polymer solutions~\cite{Gennes1979, Rubinstein2003, Wang2017}.
Using $\phi^*$ and concentration scales associated with colloidal jamming, the mechanical equation-of-state for ABPs can be collapsed onto a single curve which contains a number of previously overlooked features, the inclusion of which qualitatively alters previous predictions of the phase behavior of ABPs.
Moreover, the use of a dynamic overlap concentration scale may prove to be of great utility in the construction of accurate nonequilibrium equations-of-state.

\emph{Simulation Model and the EoS.--} In an effort to unpack the concentration dependence of the pressure, we focus on the simplest and most widely studied active system: purely-repulsive active Brownian disks.
By introducing a small degree of polydispersity in disk size, we ensure the system remains disordered for all concentrations.
This is in contrast to monodisperse active disks, which exhibit a order-disorder phase transition with a complex dependence on activity~\cite{Digregorio2018, Klamser2018, Paliwal2020}.
The small degree of polydispersity allows us to isolate the disordered branch of the equations-of-state, which not only will serve as a useful reference system for active fluids with more complicated interactions, but allows for the quantitative study of the interplay between activity and concentration.

We consider a two dimensional system of overdamped disks at fixed number density $n$ where each particle experiences a drag force $\zeta \bm{U}$, a conservative interparticle force $\bm{F^c}$ and an active self-propelling force $\bm{F^{a}}=\zeta U_0 \bm{q}$. 
The orientation of each particle $\bm{q}$ is independent and obeys diffusive rotary dynamics with a characteristic reorientation time $\tau_R$.
The sum of the forces results in a simple equation-of-motion for the particle velocity:
\begin{equation}
\label{eq:eom}
\bm{U} = (\bm{F^{a}} + \bm{F^{c}})/\zeta. 
\end{equation}
The interparticle force $\bm{F^c}$ arises from a Weeks-Chandler-Anderson (WCA) potential~\cite{Weeks2011} characterized by a potential depth $\varepsilon$ and an average Lennard-Jones diameter $d$. Particle diameters are drawn from a normal distribution with a standard deviation of $0.1 d$. 
Importantly, as our active force is of finite amplitude, a sufficiently strong choice for the repulsive force $\bm{F^{c}}$ will mimic a true hard-particle potential.
A choice of $\varepsilon/(\zeta U_0 d)=100$ is found to result in hard disk statistics with an effective average particle diameter of $2^{1/6} d$.
In this hard-disk limit, the state of our system is independent of the amplitude of the active force and is fully described by two geometric parameters: the area fraction $\phi = n\pi (2^{1/6} d)^2/4$ and the dimensionless intrinsic run-length $\ell_0 / d$.
Using the GPU-enabled HOOMD-blue software package~\cite{Anderson2020}, all simulations were conducted with 40,000 particles and run for a minimum duration of 5,000~$d/U_0$.

\begin{figure*}[t!]
	\centering
	\includegraphics[width=1\textwidth]{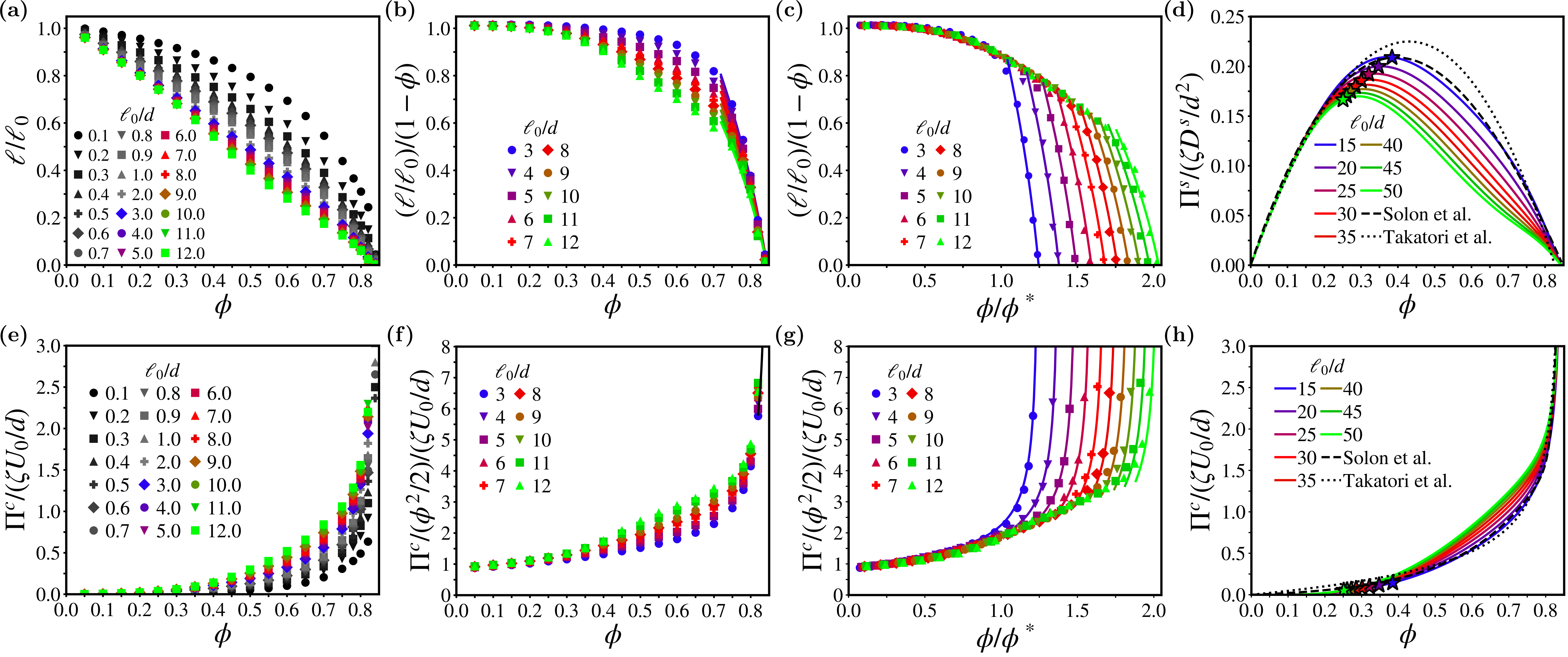}
	\caption{\protect\small{{Concentration dependence of the normalized effective run-length (a) and collisional pressure (e). The first order corrections to the normalized run-length (b) and collisional pressure (f) in the dilute limit and the asymptotic scaling near maximal-packing. Collapse of the dilute and semidilute regions for the effective run-length (c) and collisional pressure (g) using the dynamic overlap concentration $\phi^*$. Lines represent our theoretical equations-of-state in the active jamming region. Scaling-informed predicted (d) swim (normalized by the ideal swim energy scale $\zeta D^s = \zeta U_0 \ell_0/2$) and (h) collisional pressures in the region of instability. The previously assumed forms of the swim pressure used by Takatori and Brady~\cite{Takatori2015} and Solon et al.~\cite{Solon2018} are shown for comparison. For each activity curve, the identically colored star represents the location of $\phi^*$.}}}
	\label{figure2}
\end{figure*}

The total mechanical pressure $\Pi$ has two contributions: the collisional pressure $\Pi^c$ arising from conservative interparticle interactions and the swim pressure $\Pi^s$ generated by the active force. 
The collisional pressure follows from the standard micromechanical virial for conservative interactions $\Pi^c = n \langle \bm{x} \cdot \bm{F^c} \rangle/2$ where $\bm{x}$ is the particle position and $\langle ... \rangle$ denotes an average over all particles.
At steady-state, $\Pi^s$ can be written in the impulse form~\cite{Solon2018,Patch2017,Das2019}:
\begin{equation}
\label{eq:swimpressure}
\Pi^s = n \frac{\zeta U_0 \tau_R}{2} \langle \bm{q} \cdot \bm{U}\rangle.     
\end{equation}
For an isotropic system of active particles free of aligning interaction, Eq.~\eqref{eq:swimpressure}  is the pressure that the suspension would exert on a flat torque-free macroscopic boundary. 
Under these conditions, the total pressure exerted on the surroundings is simply the sum of the collisional and swim pressure $\Pi = \Pi^c + \Pi^s$.

Equation~\eqref{eq:swimpressure} allows us to directly probe via simulation the effective run-length of the particles, $\ell \equiv \tau_R \langle \bm{q} \cdot \bm{U} \rangle$, which is the true correlation length of a particle's trajectory and a measure of the correlation between the orientation of a particle and the forces (both active and interparticle) acting upon it. 
Analytical expressions for the effective run-length $\ell$ require solving a many-body dynamics problem in both position and orientation space~\cite{Bickmann2020a,Bickmann2020,Jeggle2020}, and have been derived in only a few asymptotic limits. 
Additionally, the nature of $\ell$ is further obfuscated as direct measurement is limited to a narrow region of ($\ell_0$, $\phi$)-state-space where motility-induced phase separation is absent (i.e.,~outside of the MIPS phase envelope).
Paradoxically, the prediction of both the spinodal and binodal requires knowledge of $\ell$ in regions where it cannot be directly measured.
Extrapolation of $\ell$ into these mechanically forbidden regions requires a complete understanding of the trends in $\ell$ as the system approaches the critical $\ell_0$ for MIPS $(\ell_0/d \approx 13)$. 
Figure~\ref{figure2}(a) presents the complete $\phi$ dependence of $\ell$ in regions of ($\ell_0$, $\phi$)-state-space where the system is homogeneous.

Our approach for isolating these trends is motivated by two simple observations. 
First, $\ell$ becomes increasingly independent of the intrinsic run-length $\ell_0$ with increasing concentration. 
In the dilute limit, $\ell = \ell_0 =  U_0 \tau_R$, while in the limit of maximal-packing $\ell \approx 0$.
Second, deviations from the dilute scaling prediction of $\ell$ occur at lower concentrations as $\ell_0$ is increased. 
Microrheology calculations for hard-disks show the first-order correction to the effective run-length is given by $\ell \approx \ell_0 (1-\phi)$ in the limit of large $\ell_0$~\cite{Takatori2014, Squires2005}.
As shown in Fig.~\ref{figure2}(b), this expression is in excellent agreement with our data for persistent ABPs ($\ell_0 > 3d$), and note the previously mentioned trend in the deviation from this dilute scaling as $\ell_0$ increases. 

\emph{Dynamic Overlap Concentration.--}~{The dependence of the swim pressure on a single-particle length scale at low concentrations -- the intrinsic run-length $\ell_0$ -- and its independence from that length scale at high concentrations suggests a connection to equilibrium polymer solutions.
There exist distinct concentration regimes for polymer solutions, corresponding to isolated (dilute) and overlapping chains (semidilute)~\cite{Gennes1979, Rubinstein2003, Wang2017}.
The thermodynamic equations-of-state in each regime have distinct dependencies on single-chain properties.
For example, the osmotic pressure of a dilute polymer solution is inversely proportional to the degree of polymerization $N$ [see Fig.~\ref{figure1}]. 
In the semidilute regime where chains strongly overlap, the individuality of each chain is lost resulting in an osmotic pressure that is independent of the single-chain property $N$.
The overlap concentration $\phi^*$ delineates the dilute and semidilute regimes and is the concentration at which polymer coils begin to overlap. 
The density of chains at $\phi^*$ is inversely proportional to the effective volume occupied by a single polymer chain $V^\text{chain}$.}

For ABPs, we seek the characteristic single-particle volume (area in 2D) $V^{\text{ABP}}$ that plays the analogous role to the single-chain volume $V^\text{chain}$ for polymer solutions.
In 2D, a natural starting point is to consider the area swept out by an ideal ABP in a single reorientation time $\tau_R$:  $V^{\text{ABP}} \sim \ell_0 d$.
For concentrations beyond $\phi^* \sim 1/V^{\text{ABP}}$ particle trajectories overlap and collisions prevent ABPs from executing their intrinsic random walk of length $\ell_0$ [see Fig.~\ref{figure1}].
The single-particle volume given by $V^{\text{ABP}} \sim \ell_0 d$ is likely an overestimate of the overlap probability. 
More realistically, we anticipate a weaker dependence on $\ell_0$ and introduce the more general definition $V^{\text{ABP}} \sim \ell_0^{\lambda} d^{2 - \lambda}$ where $0 < \lambda < 1$ is a constant to be determined.
The resulting overlap concentration $\phi^*$ can be written as:
\begin{equation}
\label{eq:overlap}
    \phi^* \sim \left (\frac{d}{\ell_0} \right)^{\lambda}.
\end{equation}

The dynamic overlap concentration $\phi^*$ proposed in Eq.~\eqref{eq:overlap} should delineate dilute and semidilute concentration regimes for all values of $\ell_0$. 
Using this criterion as a guide (see Supplemental Material~\cite{Note2}), we identify a significant dependence of the dynamic overlap concentration on the intrinsic run-length with $\lambda \approx 0.353$ for our polydisperse system of active Brownian disks.
As shown in Fig.~\ref{figure2}(c), rescaling $\phi$ with $\phi^*$ allows for a collapse of $\ell$ in the dilute ($\phi/\phi^* < 1$) and semidilute ($\phi/\phi^* \ge 1$) regimes. 
Notably, for each $\ell_0$ deviations from the collapse begin to occur at a $\phi^*$-independent area fraction $\phi \approx 0.72$, indicative of an upper-bound for our semidilute regime and the emergence of what we term as the active jamming regime.  

As the system transitions to the active jamming regime, the run-length tends to zero as the maximal-packing $\phi_m \approx 0.845$ is approached.
Furthermore, the run-length becomes increasingly independent of $\ell_0$ as geometric packing constraints set the scale for particle motion.
We find $\ell$ in the active jamming regime is well-described by: 
\begin{equation}
\label{eq:dense}
\ell/\ell_0=\mathcal{D}(1-e^{-\mathcal{E}/\ell_0})(1-\phi/\phi_m),
\end{equation}
where $\mathcal{D},\mathcal{E}>0$ are constant empirical fitting parameters~\cite{Note2}.
As shown by the family of curves in Fig.~\ref{figure2}(c), Eq.~\eqref{eq:dense} accurately captures the deviation of $\ell$ from the semidilute regime as a function of $\ell_0$ while also capturing our physical asymptotic expectations. With increasing $\ell_0$, the effective run-length approaches a constant value set by the packing geometry i.e.,$~\lim_{\ell_0 \to \infty} \ell = \mathcal{D}\mathcal{E}(1-\phi/\phi_m)$. Previous EoS~\cite{Takatori2015, Solon2018} which expressed $\ell = \ell_0 f(\phi)$ (for large $\ell_0$) fail to capture this physical expectation and, in fact, find the unphysical result of $\lim_{\ell_0 \to \infty} \ell = \infty$ for all $\phi$ unless $f(\phi)$ is identically zero.

\begin{figure}[t!]
	\centering
	\includegraphics[width=.475\textwidth]{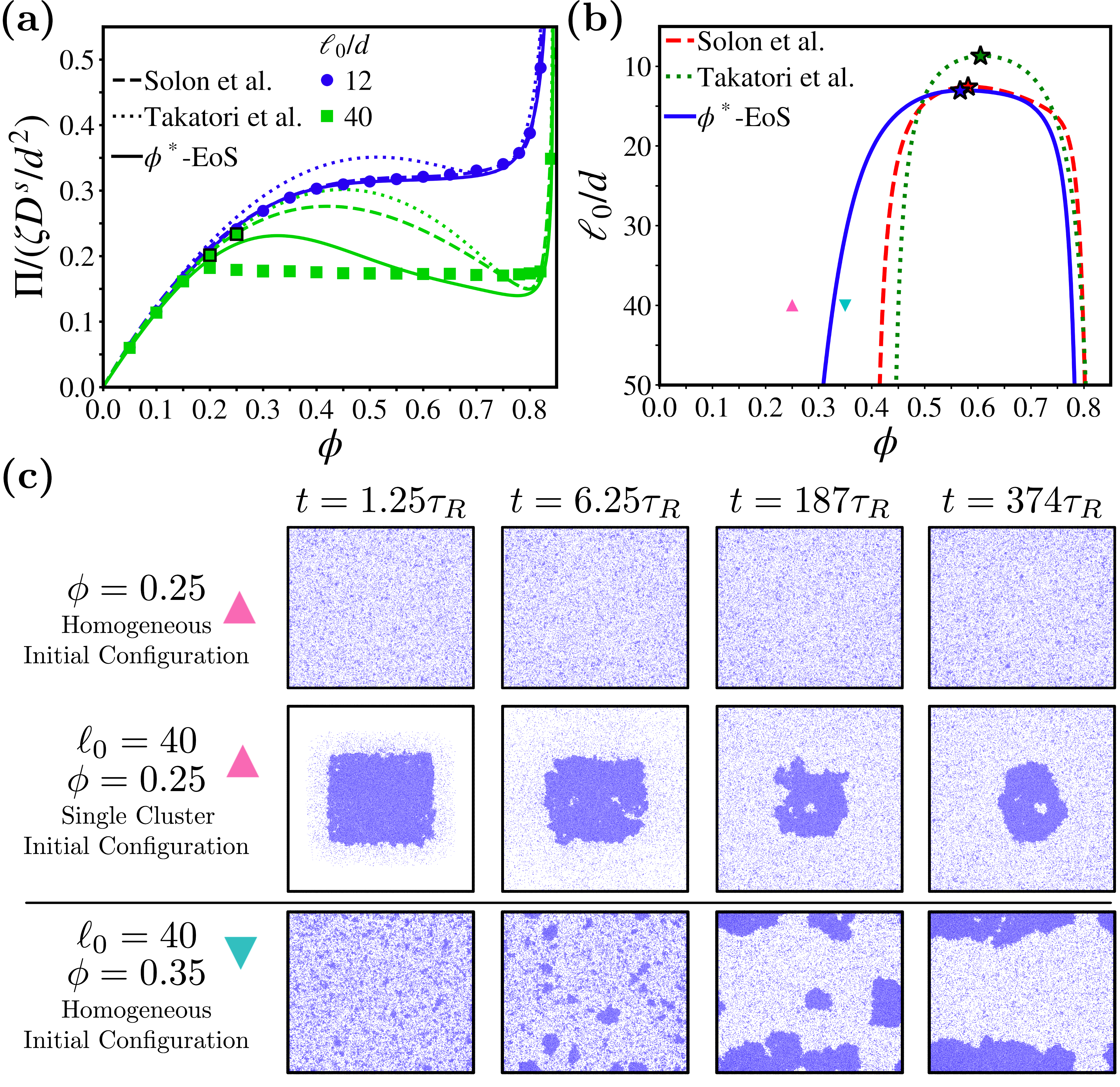}
	\caption{\protect\small{{(a) $\phi^*$-EoS in both the regions of stability and instability (metastable points are outlined in black) and (b) the predicted spinodal. The $\phi^*$-EoS and resulting spinodal are compared to those of Takatori et al.~\cite{Takatori2015} and Solon et al.~\cite{Solon2018}. Phase separation kinetics (c) confirm that our predicted spinodal indeed delineates regions of the phase diagram in which large wavelength density homogeneity is unstable from regions in which homogeneity is metastable.}}}
	\label{figure3}
\end{figure}

By identifying $\phi^*$ and the scaling behavior in the active jamming regime, the $\ell_0$ dependence of the three regions of the swim pressure (dilute, semidilute and active jamming) can be described quantitatively. 
From this information, it is straight-forward to construct a functional form that captures the full $\phi$ and $\ell_0$ dependencies of the swim pressure~\cite{Note2}.
In Fig.~\ref{figure2}(d), representative curves for the $\phi^*$-informed swim pressure are provided for a range of $\ell_0$ above the critical intrinsic run-length for MIPS. 
For comparison, we include expressions for the swim pressure put forward in previous works~\cite{Takatori2015, Solon2018}.
Importantly and in contrast to Refs.~\cite{Solon2018,Takatori2015}, the swim pressure derived using $\phi^*$ does not asymptote to a characteristic functional form at large $\ell_0$.
In fact, the dependence on $\ell_0$ is quite striking but not unexpected given our newly identified concentration regimes.

The concentration regimes defined above not only describe the swim pressure, but also reveal the $\ell_0$ and $\phi$ dependencies of the collisional pressure $\Pi^c$, as shown in Figures~\ref{figure2}(e)-(h) (see \cite{Note2} for further discussion).
While existing works~\cite{Takatori2015, Solon2018} have proposed that $\Pi^c$ is independent of $\ell_0$, a clear dependence can be observed in Fig.~\ref{figure2}(f) and collapsed using $\phi^*$ [see Fig.~\ref{figure2}(g)]. 
Deviations from the collapse again occur in the active jamming region, where $\Pi^c$ is entirely independent of $\ell_0$ and is well-described by an EoS for \textit{passive} polydisperse disks~\cite{Santos2017} (with $\zeta U_0$ replacing the thermal force scale $k_BT/d$). 

\emph{Phase Behavior Implications.--} 
In Fig.~\ref{figure3}(a), we compare the total pressure predicted by our dynamic overlap equation of state ($\phi^*$-EoS) with those derived in previous work.
We consider two values of $\ell_0$, one near the critical intrinsic run-length $\ell_0/d=12$ and one deep within the MIPS coexistence region $\ell_0/d=40$.
At $\ell_0/d=12$, the three expressions are found to be in relatively good agreement, while for $\ell_0/d=40$ there is a substantial disagreement between the $\phi^*$-EoS and those proposed in previous works, 
most notably in the range of concentrations where the equation of state becomes unphysical (manifested as a van der Waals loop) and exhibits a mechanical instability (i.e. $(\partial \Pi/ \partial \phi)_{\ell_0} < 0$).
The bounds (the stability limit) of this mechanically unstable region denote the spinodal, shown in full in Fig.~\ref{figure3}(b). 

Compared to previous works, the $\phi^*$-EoS predicts a much broader unstable region -- a prediction that can be directly tested via simulation. 
In Fig.~\ref{figure3}(c), we show a series of snapshots that illustrate the phase separation kinetics for different points in the stability diagram. 
For the ($\ell_0/d$, $\phi$)-state-space point $(40,0.35)$, all simulations, independent of the initial configuration, rapidly phase separate  [see bottom row of Fig.~\ref{figure3}(c)] -- an indication that this state-space point is unstable (within the spinodal).
While, for the ($\ell_0/d$, $\phi$)-state-space point $(40,0.25)$, it is possible to stabilize both a homogeneous and phase-separated configuration depending on the choice of initial configuration [see top two rows of Fig.~\ref{figure3}(c)].
The metastability at this state-space point demonstrates that it is outside of the spinodal but still within the binodal for MIPS.
That these two state points of contrasting stability straddle our predicted spinodal provides an independent verification of the accuracy of $\phi^*$-EoS well-beyond the critical point.
In addition, the $\phi^*$-EoS captures features that are central for correctly \textit{predicting} the binodal for MIPS, which will be detailed in a future publication.

\emph{Conclusions.--} 
We close by emphasizing that the unique features of the EoS identified in this work were \textit{revealed} through the recognition of the existence of a dynamic overlap concentration scale. 
It is our hope and expectation that the use of trajectory-space length scales to define concentration scales will prove to be of utility in the study of non-equilibrium systems. 
In future work, it will be interesting to unravel the dependencies of the dynamic overlap concentration on the nature of the particle interactions and dynamics as well as the system dimensionality.

\begin{acknowledgments}
\emph{Acknowledgments.--}
S.A.M. acknowledges financial support from the Arnold and Mabel Beckman Foundation. A.K.O. acknowledges support from the Schmidt Science Fellowship in partnership with the Rhodes Trust. J.F.B. acknowledges support by the National Science Foundation under Grant No. CBET-1803662.  We gratefully acknowledge the support of the NVIDIA Corporation for the donation of the Titan V GPU used to carry out this work.
\end{acknowledgments}


\begin{thebibliography}{66}%
\makeatletter
\providecommand \@ifxundefined [1]{%
 \@ifx{#1\undefined}
}%
\providecommand \@ifnum [1]{%
 \ifnum #1\expandafter \@firstoftwo
 \else \expandafter \@secondoftwo
 \fi
}%
\providecommand \@ifx [1]{%
 \ifx #1\expandafter \@firstoftwo
 \else \expandafter \@secondoftwo
 \fi
}%
\providecommand \natexlab [1]{#1}%
\providecommand \enquote  [1]{``#1''}%
\providecommand \bibnamefont  [1]{#1}%
\providecommand \bibfnamefont [1]{#1}%
\providecommand \citenamefont [1]{#1}%
\providecommand \href@noop [0]{\@secondoftwo}%
\providecommand \href [0]{\begingroup \@sanitize@url \@href}%
\providecommand \@href[1]{\@@startlink{#1}\@@href}%
\providecommand \@@href[1]{\endgroup#1\@@endlink}%
\providecommand \@sanitize@url [0]{\catcode `\\12\catcode `\$12\catcode
  `\&12\catcode `\#12\catcode `\^12\catcode `\_12\catcode `\%12\relax}%
\providecommand \@@startlink[1]{}%
\providecommand \@@endlink[0]{}%
\providecommand \url  [0]{\begingroup\@sanitize@url \@url }%
\providecommand \@url [1]{\endgroup\@href {#1}{\urlprefix }}%
\providecommand \urlprefix  [0]{URL }%
\providecommand \Eprint [0]{\href }%
\providecommand \doibase [0]{https://doi.org/}%
\providecommand \selectlanguage [0]{\@gobble}%
\providecommand \bibinfo  [0]{\@secondoftwo}%
\providecommand \bibfield  [0]{\@secondoftwo}%
\providecommand \translation [1]{[#1]}%
\providecommand \BibitemOpen [0]{}%
\providecommand \bibitemStop [0]{}%
\providecommand \bibitemNoStop [0]{.\EOS\space}%
\providecommand \EOS [0]{\spacefactor3000\relax}%
\providecommand \BibitemShut  [1]{\csname bibitem#1\endcsname}%
\let\auto@bib@innerbib\@empty
\bibitem [{\citenamefont {Fily}\ \emph {et~al.}(2014)\citenamefont {Fily},
  \citenamefont {Henkes},\ and\ \citenamefont {Marchetti}}]{Fily2014a}%
  \BibitemOpen
  \bibfield  {author} {\bibinfo {author} {\bibfnamefont {Y.}~\bibnamefont
  {Fily}}, \bibinfo {author} {\bibfnamefont {S.}~\bibnamefont {Henkes}},\ and\
  \bibinfo {author} {\bibfnamefont {M.~C.}\ \bibnamefont {Marchetti}},\ }\href
  {https://doi.org/10.1039/c3sm52469h} {\bibfield  {journal} {\bibinfo
  {journal} {Soft Matter}\ }\textbf {\bibinfo {volume} {10}},\ \bibinfo {pages}
  {2132} (\bibinfo {year} {2014})}\BibitemShut {NoStop}%
\bibitem [{\citenamefont {Mallory}\ \emph {et~al.}(2014)\citenamefont
  {Mallory}, \citenamefont {{\v{S}}ari{\'{c}}}, \citenamefont {Valeriani},\
  and\ \citenamefont {Cacciuto}}]{Mallory2014b}%
  \BibitemOpen
  \bibfield  {author} {\bibinfo {author} {\bibfnamefont {S.~A.}\ \bibnamefont
  {Mallory}}, \bibinfo {author} {\bibfnamefont {A.}~\bibnamefont
  {{\v{S}}ari{\'{c}}}}, \bibinfo {author} {\bibfnamefont {C.}~\bibnamefont
  {Valeriani}},\ and\ \bibinfo {author} {\bibfnamefont {A.}~\bibnamefont
  {Cacciuto}},\ }\href {https://doi.org/10.1103/PhysRevE.89.052303} {\bibfield
  {journal} {\bibinfo  {journal} {Phys. Rev. E}\ }\textbf {\bibinfo {volume}
  {89}},\ \bibinfo {pages} {52303} (\bibinfo {year} {2014})}\BibitemShut
  {NoStop}%
\bibitem [{\citenamefont {Takatori}\ \emph {et~al.}(2014)\citenamefont
  {Takatori}, \citenamefont {Yan},\ and\ \citenamefont {Brady}}]{Takatori2014}%
  \BibitemOpen
  \bibfield  {author} {\bibinfo {author} {\bibfnamefont {S.~C.}\ \bibnamefont
  {Takatori}}, \bibinfo {author} {\bibfnamefont {W.}~\bibnamefont {Yan}},\ and\
  \bibinfo {author} {\bibfnamefont {J.~F.}\ \bibnamefont {Brady}},\ }\href
  {https://doi.org/10.1103/PhysRevLett.113.028103} {\bibfield  {journal}
  {\bibinfo  {journal} {Phys. Rev. Lett.}\ }\textbf {\bibinfo {volume} {113}},\
  \bibinfo {pages} {28103} (\bibinfo {year} {2014})}\BibitemShut {NoStop}%
\bibitem [{\citenamefont {Solon}\ \emph
  {et~al.}(2015{\natexlab{a}})\citenamefont {Solon}, \citenamefont
  {Stenhammar}, \citenamefont {Wittkowski}, \citenamefont {Kardar},
  \citenamefont {Kafri}, \citenamefont {Cates},\ and\ \citenamefont
  {Tailleur}}]{Solon2015}%
  \BibitemOpen
  \bibfield  {author} {\bibinfo {author} {\bibfnamefont {A.~P.}\ \bibnamefont
  {Solon}}, \bibinfo {author} {\bibfnamefont {J.}~\bibnamefont {Stenhammar}},
  \bibinfo {author} {\bibfnamefont {R.}~\bibnamefont {Wittkowski}}, \bibinfo
  {author} {\bibfnamefont {M.}~\bibnamefont {Kardar}}, \bibinfo {author}
  {\bibfnamefont {Y.}~\bibnamefont {Kafri}}, \bibinfo {author} {\bibfnamefont
  {M.~E.}\ \bibnamefont {Cates}},\ and\ \bibinfo {author} {\bibfnamefont
  {J.}~\bibnamefont {Tailleur}},\ }\href
  {https://doi.org/10.1103/PhysRevLett.114.198301} {\bibfield  {journal}
  {\bibinfo  {journal} {Phys. Rev. Lett.}\ }\textbf {\bibinfo {volume} {114}},\
  \bibinfo {pages} {198301} (\bibinfo {year} {2015}{\natexlab{a}})}\BibitemShut
  {NoStop}%
\bibitem [{\citenamefont {Solon}\ \emph
  {et~al.}(2015{\natexlab{b}})\citenamefont {Solon}, \citenamefont {Fily},
  \citenamefont {Baskaran}, \citenamefont {Cates}, \citenamefont {Kafri},
  \citenamefont {Kardar},\ and\ \citenamefont
  {Tailleur}}]{solon_pressure_2015-1}%
  \BibitemOpen
  \bibfield  {author} {\bibinfo {author} {\bibfnamefont {A.~P.}\ \bibnamefont
  {Solon}}, \bibinfo {author} {\bibfnamefont {Y.}~\bibnamefont {Fily}},
  \bibinfo {author} {\bibfnamefont {A.}~\bibnamefont {Baskaran}}, \bibinfo
  {author} {\bibfnamefont {M.~E.}\ \bibnamefont {Cates}}, \bibinfo {author}
  {\bibfnamefont {Y.}~\bibnamefont {Kafri}}, \bibinfo {author} {\bibfnamefont
  {M.}~\bibnamefont {Kardar}},\ and\ \bibinfo {author} {\bibfnamefont
  {J.}~\bibnamefont {Tailleur}},\ }\href {https://doi.org/10.1038/nphys3377}
  {\bibfield  {journal} {\bibinfo  {journal} {Nat. Phys.}\ }\textbf {\bibinfo
  {volume} {11}},\ \bibinfo {pages} {673} (\bibinfo {year}
  {2015}{\natexlab{b}})}\BibitemShut {NoStop}%
\bibitem [{\citenamefont {Smallenburg}\ and\ \citenamefont
  {L{\"{o}}wen}(2015)}]{Smallenburg2015}%
  \BibitemOpen
  \bibfield  {author} {\bibinfo {author} {\bibfnamefont {F.}~\bibnamefont
  {Smallenburg}}\ and\ \bibinfo {author} {\bibfnamefont {H.}~\bibnamefont
  {L{\"{o}}wen}},\ }\href {https://doi.org/10.1103/PhysRevE.92.032304}
  {\bibfield  {journal} {\bibinfo  {journal} {Phys. Rev. E}\ }\textbf {\bibinfo
  {volume} {92}},\ \bibinfo {pages} {32304} (\bibinfo {year}
  {2015})}\BibitemShut {NoStop}%
\bibitem [{\citenamefont {Winkler}\ \emph {et~al.}(2015)\citenamefont
  {Winkler}, \citenamefont {Wysocki},\ and\ \citenamefont
  {Gompper}}]{Winkler2015}%
  \BibitemOpen
  \bibfield  {author} {\bibinfo {author} {\bibfnamefont {R.~G.}\ \bibnamefont
  {Winkler}}, \bibinfo {author} {\bibfnamefont {A.}~\bibnamefont {Wysocki}},\
  and\ \bibinfo {author} {\bibfnamefont {G.}~\bibnamefont {Gompper}},\ }\href
  {https://doi.org/10.1039/c5sm01412c} {\bibfield  {journal} {\bibinfo
  {journal} {Soft Matter}\ }\textbf {\bibinfo {volume} {11}},\ \bibinfo {pages}
  {6680} (\bibinfo {year} {2015})}\BibitemShut {NoStop}%
\bibitem [{\citenamefont {Speck}\ and\ \citenamefont
  {Jack}(2016)}]{speck_ideal_2016}%
  \BibitemOpen
  \bibfield  {author} {\bibinfo {author} {\bibfnamefont {T.}~\bibnamefont
  {Speck}}\ and\ \bibinfo {author} {\bibfnamefont {R.~L.}\ \bibnamefont
  {Jack}},\ }\href {https://doi.org/10.1103/PhysRevE.93.062605} {\bibfield
  {journal} {\bibinfo  {journal} {Phys. Rev. E}\ }\textbf {\bibinfo {volume}
  {93}},\ \bibinfo {pages} {62605} (\bibinfo {year} {2016})}\BibitemShut
  {NoStop}%
\bibitem [{\citenamefont {Joyeux}\ and\ \citenamefont
  {Bertin}(2016)}]{Joyeux2016}%
  \BibitemOpen
  \bibfield  {author} {\bibinfo {author} {\bibfnamefont {M.}~\bibnamefont
  {Joyeux}}\ and\ \bibinfo {author} {\bibfnamefont {E.}~\bibnamefont
  {Bertin}},\ }\href {https://doi.org/10.1103/PhysRevE.93.032605} {\bibfield
  {journal} {\bibinfo  {journal} {Phys. Rev. E}\ }\textbf {\bibinfo {volume}
  {93}},\ \bibinfo {pages} {032605} (\bibinfo {year} {2016})}\BibitemShut
  {NoStop}%
\bibitem [{\citenamefont {Fily}\ \emph {et~al.}(2018)\citenamefont {Fily},
  \citenamefont {Kafri}, \citenamefont {Solon}, \citenamefont {Tailleur},\ and\
  \citenamefont {Turner}}]{fily2018}%
  \BibitemOpen
  \bibfield  {author} {\bibinfo {author} {\bibfnamefont {Y.}~\bibnamefont
  {Fily}}, \bibinfo {author} {\bibfnamefont {Y.}~\bibnamefont {Kafri}},
  \bibinfo {author} {\bibfnamefont {A.~P.}\ \bibnamefont {Solon}}, \bibinfo
  {author} {\bibfnamefont {J.}~\bibnamefont {Tailleur}},\ and\ \bibinfo
  {author} {\bibfnamefont {A.}~\bibnamefont {Turner}},\ }\href
  {https://doi.org/10.1088/1751-8121/aa99b6} {\bibfield  {journal} {\bibinfo
  {journal} {J. Phys. A Math. Theor.}\ }\textbf {\bibinfo {volume} {51}},\
  \bibinfo {pages} {044003} (\bibinfo {year} {2018})}\BibitemShut {NoStop}%
\bibitem [{\citenamefont {Marconi}\ \emph
  {et~al.}(2017{\natexlab{a}})\citenamefont {Marconi}, \citenamefont {Maggi},\
  and\ \citenamefont {Paoluzzi}}]{MariniBettoloMarconi2017}%
  \BibitemOpen
  \bibfield  {author} {\bibinfo {author} {\bibfnamefont {U.~M.~B.}\
  \bibnamefont {Marconi}}, \bibinfo {author} {\bibfnamefont {C.}~\bibnamefont
  {Maggi}},\ and\ \bibinfo {author} {\bibfnamefont {M.}~\bibnamefont
  {Paoluzzi}},\ }\href {https://doi.org/10.1063/1.4991731} {\bibfield
  {journal} {\bibinfo  {journal} {J. Chem. Phys.}\ }\textbf {\bibinfo {volume}
  {147}},\ \bibinfo {pages} {024903} (\bibinfo {year}
  {2017}{\natexlab{a}})}\BibitemShut {NoStop}%
\bibitem [{\citenamefont {Patch}\ \emph {et~al.}(2017)\citenamefont {Patch},
  \citenamefont {Yllanes},\ and\ \citenamefont {Marchetti}}]{Patch2017}%
  \BibitemOpen
  \bibfield  {author} {\bibinfo {author} {\bibfnamefont {A.}~\bibnamefont
  {Patch}}, \bibinfo {author} {\bibfnamefont {D.}~\bibnamefont {Yllanes}},\
  and\ \bibinfo {author} {\bibfnamefont {M.~C.}\ \bibnamefont {Marchetti}},\
  }\href {https://doi.org/10.1103/PhysRevE.95.012601} {\bibfield  {journal}
  {\bibinfo  {journal} {Phys. Rev. E}\ }\textbf {\bibinfo {volume} {95}},\
  \bibinfo {pages} {012601} (\bibinfo {year} {2017})}\BibitemShut {NoStop}%
\bibitem [{\citenamefont {Das}\ \emph {et~al.}(2019)\citenamefont {Das},
  \citenamefont {Gompper},\ and\ \citenamefont {Winkler}}]{Das2019}%
  \BibitemOpen
  \bibfield  {author} {\bibinfo {author} {\bibfnamefont {S.}~\bibnamefont
  {Das}}, \bibinfo {author} {\bibfnamefont {G.}~\bibnamefont {Gompper}},\ and\
  \bibinfo {author} {\bibfnamefont {R.~G.}\ \bibnamefont {Winkler}},\ }\href
  {http://arxiv.org/abs/1902.07435} {\bibfield  {journal} {\bibinfo  {journal}
  {Sci. Rep.}\ }\textbf {\bibinfo {volume} {9}} (\bibinfo {year}
  {2019})}\BibitemShut {NoStop}%
\bibitem [{\citenamefont {Epstein}\ \emph {et~al.}(2019)\citenamefont
  {Epstein}, \citenamefont {Klymko},\ and\ \citenamefont
  {Mandadapu}}]{Epstein2019}%
  \BibitemOpen
  \bibfield  {author} {\bibinfo {author} {\bibfnamefont {J.~M.}\ \bibnamefont
  {Epstein}}, \bibinfo {author} {\bibfnamefont {K.}~\bibnamefont {Klymko}},\
  and\ \bibinfo {author} {\bibfnamefont {K.~K.}\ \bibnamefont {Mandadapu}},\
  }\href {https://doi.org/10.1063/1.5054912} {\bibfield  {journal} {\bibinfo
  {journal} {J. Chem. Phys.}\ }\textbf {\bibinfo {volume} {150}},\ \bibinfo
  {pages} {194109} (\bibinfo {year} {2019})}\BibitemShut {NoStop}%
\bibitem [{\citenamefont {Omar}\ \emph {et~al.}(2020)\citenamefont {Omar},
  \citenamefont {Wang},\ and\ \citenamefont {Brady}}]{Omar2020}%
  \BibitemOpen
  \bibfield  {author} {\bibinfo {author} {\bibfnamefont {A.~K.}\ \bibnamefont
  {Omar}}, \bibinfo {author} {\bibfnamefont {Z.~G.}\ \bibnamefont {Wang}},\
  and\ \bibinfo {author} {\bibfnamefont {J.~F.}\ \bibnamefont {Brady}},\ }\href
  {https://doi.org/10.1103/PhysRevE.101.012604} {\bibfield  {journal} {\bibinfo
   {journal} {Phys. Rev. E}\ }\textbf {\bibinfo {volume} {101}},\ \bibinfo
  {pages} {12604} (\bibinfo {year} {2020})}\BibitemShut {NoStop}%
\bibitem [{\citenamefont {Sokolov}\ \emph {et~al.}(2018)\citenamefont
  {Sokolov}, \citenamefont {Rubio}, \citenamefont {Brady},\ and\ \citenamefont
  {Aranson}}]{Sokolov2018}%
  \BibitemOpen
  \bibfield  {author} {\bibinfo {author} {\bibfnamefont {A.}~\bibnamefont
  {Sokolov}}, \bibinfo {author} {\bibfnamefont {L.~D.}\ \bibnamefont {Rubio}},
  \bibinfo {author} {\bibfnamefont {J.~F.}\ \bibnamefont {Brady}},\ and\
  \bibinfo {author} {\bibfnamefont {I.~S.}\ \bibnamefont {Aranson}},\ }\href
  {https://doi.org/10.1038/s41467-018-03758-z} {\bibfield  {journal} {\bibinfo
  {journal} {Nat. Commun.}\ }\textbf {\bibinfo {volume} {9}},\ \bibinfo {pages}
  {1322} (\bibinfo {year} {2018})}\BibitemShut {NoStop}%
\bibitem [{\citenamefont {Li}\ and\ \citenamefont {{Ten
  Wolde}}(2019)}]{Li2019}%
  \BibitemOpen
  \bibfield  {author} {\bibinfo {author} {\bibfnamefont {Y.}~\bibnamefont
  {Li}}\ and\ \bibinfo {author} {\bibfnamefont {P.~R.}\ \bibnamefont {{ten
  Wolde}}},\ }\href {https://doi.org/10.1103/PhysRevLett.123.148003} {\bibfield
   {journal} {\bibinfo  {journal} {Phys. Rev. Lett.}\ }\textbf {\bibinfo
  {volume} {123}},\ \bibinfo {pages} {148003} (\bibinfo {year}
  {2019})}\BibitemShut {NoStop}%
\bibitem [{\citenamefont {Paoluzzi}\ \emph {et~al.}(2016)\citenamefont
  {Paoluzzi}, \citenamefont {{Di Leonardo}}, \citenamefont {Marchetti},\ and\
  \citenamefont {Angelani}}]{Paoluzzi2016}%
  \BibitemOpen
  \bibfield  {author} {\bibinfo {author} {\bibfnamefont {M.}~\bibnamefont
  {Paoluzzi}}, \bibinfo {author} {\bibfnamefont {R.}~\bibnamefont {{Di
  Leonardo}}}, \bibinfo {author} {\bibfnamefont {M.~C.}\ \bibnamefont
  {Marchetti}},\ and\ \bibinfo {author} {\bibfnamefont {L.}~\bibnamefont
  {Angelani}},\ }\href {https://doi.org/10.1038/srep34146} {\bibfield
  {journal} {\bibinfo  {journal} {Sci. Rep.}\ }\textbf {\bibinfo {volume}
  {6}},\ \bibinfo {pages} {34146} (\bibinfo {year} {2016})}\BibitemShut
  {NoStop}%
\bibitem [{\citenamefont {Wang}\ \emph {et~al.}(2019)\citenamefont {Wang},
  \citenamefont {Guo}, \citenamefont {Tian},\ and\ \citenamefont
  {Chen}}]{Wang2019}%
  \BibitemOpen
  \bibfield  {author} {\bibinfo {author} {\bibfnamefont {C.}~\bibnamefont
  {Wang}}, \bibinfo {author} {\bibfnamefont {Y.~K.}\ \bibnamefont {Guo}},
  \bibinfo {author} {\bibfnamefont {W.~D.}\ \bibnamefont {Tian}},\ and\
  \bibinfo {author} {\bibfnamefont {K.}~\bibnamefont {Chen}},\ }\href
  {https://doi.org/10.1063/1.5078694} {\bibfield  {journal} {\bibinfo
  {journal} {J. Chem. Phys.}\ }\textbf {\bibinfo {volume} {150}},\ \bibinfo
  {pages} {044907} (\bibinfo {year} {2019})}\BibitemShut {NoStop}%
\bibitem [{\citenamefont {Takatori}\ and\ \citenamefont
  {Sahu}(2020)}]{Takatori2020}%
  \BibitemOpen
  \bibfield  {author} {\bibinfo {author} {\bibfnamefont {S.~C.}\ \bibnamefont
  {Takatori}}\ and\ \bibinfo {author} {\bibfnamefont {A.}~\bibnamefont
  {Sahu}},\ }\href {https://doi.org/10.1103/PhysRevLett.124.158102} {\bibfield
  {journal} {\bibinfo  {journal} {Phys. Rev. Lett.}\ }\textbf {\bibinfo
  {volume} {124}},\ \bibinfo {pages} {158102} (\bibinfo {year}
  {2020})}\BibitemShut {NoStop}%
\bibitem [{\citenamefont {Harder}\ \emph {et~al.}(2014)\citenamefont {Harder},
  \citenamefont {Mallory}, \citenamefont {Tung}, \citenamefont {Valeriani},\
  and\ \citenamefont {Cacciuto}}]{Harder2014}%
  \BibitemOpen
  \bibfield  {author} {\bibinfo {author} {\bibfnamefont {J.}~\bibnamefont
  {Harder}}, \bibinfo {author} {\bibfnamefont {S.~A.}\ \bibnamefont {Mallory}},
  \bibinfo {author} {\bibfnamefont {C.}~\bibnamefont {Tung}}, \bibinfo {author}
  {\bibfnamefont {C.}~\bibnamefont {Valeriani}},\ and\ \bibinfo {author}
  {\bibfnamefont {A.}~\bibnamefont {Cacciuto}},\ }\href
  {https://doi.org/10.1063/1.4900720} {\bibfield  {journal} {\bibinfo
  {journal} {J. Chem. Phys.}\ }\textbf {\bibinfo {volume} {141}},\ \bibinfo
  {pages} {194901} (\bibinfo {year} {2014})}\BibitemShut {NoStop}%
\bibitem [{\citenamefont {{Zaeifi Yamchi}}\ and\ \citenamefont
  {Naji}(2017)}]{ZaeifiYamchi2017}%
  \BibitemOpen
  \bibfield  {author} {\bibinfo {author} {\bibfnamefont {M.}~\bibnamefont
  {{Zaeifi Yamchi}}}\ and\ \bibinfo {author} {\bibfnamefont {A.}~\bibnamefont
  {Naji}},\ }\href {https://doi.org/10.1063/1.5001505} {\bibfield  {journal}
  {\bibinfo  {journal} {J. Chem. Phys.}\ }\textbf {\bibinfo {volume} {147}},\
  \bibinfo {pages} {194901} (\bibinfo {year} {2017})}\BibitemShut {NoStop}%
\bibitem [{\citenamefont {Ni}\ \emph {et~al.}(2015)\citenamefont {Ni},
  \citenamefont {{Cohen Stuart}},\ and\ \citenamefont {Bolhuis}}]{Ni2015}%
  \BibitemOpen
  \bibfield  {author} {\bibinfo {author} {\bibfnamefont {R.}~\bibnamefont
  {Ni}}, \bibinfo {author} {\bibfnamefont {M.~A.}\ \bibnamefont {{Cohen
  Stuart}}},\ and\ \bibinfo {author} {\bibfnamefont {P.~G.}\ \bibnamefont
  {Bolhuis}},\ }\href {https://doi.org/10.1103/PhysRevLett.114.018302}
  {\bibfield  {journal} {\bibinfo  {journal} {Phys. Rev. Lett.}\ }\textbf
  {\bibinfo {volume} {114}},\ \bibinfo {pages} {18302} (\bibinfo {year}
  {2015})}\BibitemShut {NoStop}%
\bibitem [{\citenamefont {Mallory}\ \emph {et~al.}(2020)\citenamefont
  {Mallory}, \citenamefont {Bowers},\ and\ \citenamefont
  {Cacciuto}}]{Mallory2020}%
  \BibitemOpen
  \bibfield  {author} {\bibinfo {author} {\bibfnamefont {S.~A.}\ \bibnamefont
  {Mallory}}, \bibinfo {author} {\bibfnamefont {M.~L.}\ \bibnamefont
  {Bowers}},\ and\ \bibinfo {author} {\bibfnamefont {A.}~\bibnamefont
  {Cacciuto}},\ }\href {https://doi.org/10.1063/5.0016514} {\bibfield
  {journal} {\bibinfo  {journal} {J. Chem. Phys.}\ }\textbf {\bibinfo {volume}
  {153}},\ \bibinfo {pages} {084901} (\bibinfo {year} {2020})}\BibitemShut
  {NoStop}%
\bibitem [{\citenamefont {Szakasits}\ \emph {et~al.}(2017)\citenamefont
  {Szakasits}, \citenamefont {Zhang},\ and\ \citenamefont
  {Solomon}}]{Szakasits2017}%
  \BibitemOpen
  \bibfield  {author} {\bibinfo {author} {\bibfnamefont {M.~E.}\ \bibnamefont
  {Szakasits}}, \bibinfo {author} {\bibfnamefont {W.}~\bibnamefont {Zhang}},\
  and\ \bibinfo {author} {\bibfnamefont {M.~J.}\ \bibnamefont {Solomon}},\
  }\href {https://doi.org/10.1103/PhysRevLett.119.058001} {\bibfield  {journal}
  {\bibinfo  {journal} {Phys. Rev. Lett.}\ }\textbf {\bibinfo {volume} {119}},\
  \bibinfo {pages} {58001} (\bibinfo {year} {2017})}\BibitemShut {NoStop}%
\bibitem [{\citenamefont {Omar}\ \emph {et~al.}(2019)\citenamefont {Omar},
  \citenamefont {Wu}, \citenamefont {Wang},\ and\ \citenamefont
  {Brady}}]{Omar2019}%
  \BibitemOpen
  \bibfield  {author} {\bibinfo {author} {\bibfnamefont {A.~K.}\ \bibnamefont
  {Omar}}, \bibinfo {author} {\bibfnamefont {Y.}~\bibnamefont {Wu}}, \bibinfo
  {author} {\bibfnamefont {Z.~G.}\ \bibnamefont {Wang}},\ and\ \bibinfo
  {author} {\bibfnamefont {J.~F.}\ \bibnamefont {Brady}},\ }\href
  {https://doi.org/10.1021/acsnano.8b07421} {\bibfield  {journal} {\bibinfo
  {journal} {ACS Nano}\ }\textbf {\bibinfo {volume} {13}},\ \bibinfo {pages}
  {560} (\bibinfo {year} {2019})}\BibitemShut {NoStop}%
\bibitem [{\citenamefont {Mallory}\ \emph {et~al.}(2015)\citenamefont
  {Mallory}, \citenamefont {Valeriani},\ and\ \citenamefont
  {Cacciuto}}]{Mallory2015}%
  \BibitemOpen
  \bibfield  {author} {\bibinfo {author} {\bibfnamefont {S.~A.}\ \bibnamefont
  {Mallory}}, \bibinfo {author} {\bibfnamefont {C.}~\bibnamefont {Valeriani}},\
  and\ \bibinfo {author} {\bibfnamefont {A.}~\bibnamefont {Cacciuto}},\ }\href
  {https://doi.org/10.1103/PhysRevE.92.012314} {\bibfield  {journal} {\bibinfo
  {journal} {Phys. Rev. E}\ }\textbf {\bibinfo {volume} {92}},\ \bibinfo
  {pages} {012314} (\bibinfo {year} {2015})}\BibitemShut {NoStop}%
\bibitem [{\citenamefont {Kaiser}\ and\ \citenamefont
  {L{\"{o}}wen}(2014)}]{Kaiser2014a}%
  \BibitemOpen
  \bibfield  {author} {\bibinfo {author} {\bibfnamefont {A.}~\bibnamefont
  {Kaiser}}\ and\ \bibinfo {author} {\bibfnamefont {H.}~\bibnamefont
  {L{\"{o}}wen}},\ }\href {https://doi.org/http://dx.doi.org/10.1063/1.4891095}
  {\bibfield  {journal} {\bibinfo  {journal} {J. Chem. Phys.}\ }\textbf
  {\bibinfo {volume} {141}},\  (\bibinfo {year} {2014})}\BibitemShut {NoStop}%
\bibitem [{\citenamefont {Kaiser}\ \emph {et~al.}(2015)\citenamefont {Kaiser},
  \citenamefont {Babel}, \citenamefont {{Ten Hagen}}, \citenamefont {{Von
  Ferber}},\ and\ \citenamefont {L{\"{o}}wen}}]{Kaiser2015}%
  \BibitemOpen
  \bibfield  {author} {\bibinfo {author} {\bibfnamefont {A.}~\bibnamefont
  {Kaiser}}, \bibinfo {author} {\bibfnamefont {S.}~\bibnamefont {Babel}},
  \bibinfo {author} {\bibfnamefont {B.}~\bibnamefont {{Ten Hagen}}}, \bibinfo
  {author} {\bibfnamefont {C.}~\bibnamefont {{Von Ferber}}},\ and\ \bibinfo
  {author} {\bibfnamefont {H.}~\bibnamefont {L{\"{o}}wen}},\ }\href
  {https://doi.org/10.1063/1.4916134} {\bibfield  {journal} {\bibinfo
  {journal} {J. Chem. Phys.}\ }\textbf {\bibinfo {volume} {142}},\ \bibinfo
  {pages} {124905} (\bibinfo {year} {2015})}\BibitemShut {NoStop}%
\bibitem [{\citenamefont {Xia}\ \emph {et~al.}(2019)\citenamefont {Xia},
  \citenamefont {Tian}, \citenamefont {Chen},\ and\ \citenamefont
  {Ma}}]{Xia2019}%
  \BibitemOpen
  \bibfield  {author} {\bibinfo {author} {\bibfnamefont {Y.~Q.}\ \bibnamefont
  {Xia}}, \bibinfo {author} {\bibfnamefont {W.~D.}\ \bibnamefont {Tian}},
  \bibinfo {author} {\bibfnamefont {K.}~\bibnamefont {Chen}},\ and\ \bibinfo
  {author} {\bibfnamefont {Y.~Q.}\ \bibnamefont {Ma}},\ }\href
  {https://doi.org/10.1039/c8cp05976d} {\bibfield  {journal} {\bibinfo
  {journal} {Phys. Chem. Chem. Phys.}\ }\textbf {\bibinfo {volume} {21}},\
  \bibinfo {pages} {4487} (\bibinfo {year} {2019})}\BibitemShut {NoStop}%
\bibitem [{\citenamefont {Takatori}\ and\ \citenamefont
  {Brady}(2016)}]{Takatori2016a}%
  \BibitemOpen
  \bibfield  {author} {\bibinfo {author} {\bibfnamefont {S.~C.}\ \bibnamefont
  {Takatori}}\ and\ \bibinfo {author} {\bibfnamefont {J.~F.}\ \bibnamefont
  {Brady}},\ }\href {https://doi.org/10.1016/j.cocis.2015.12.003} {\bibfield
  {journal} {\bibinfo  {journal} {Curr. Opin. Colloid Interface Sci.}\ }\textbf
  {\bibinfo {volume} {21}},\ \bibinfo {pages} {24} (\bibinfo {year}
  {2016})}\BibitemShut {NoStop}%
\bibitem [{\citenamefont {Lee}(2017)}]{Lee2017}%
  \BibitemOpen
  \bibfield  {author} {\bibinfo {author} {\bibfnamefont {C.~F.}\ \bibnamefont
  {Lee}},\ }\href {https://doi.org/10.1039/C6SM01978A} {\bibfield  {journal}
  {\bibinfo  {journal} {Soft Matter}\ }\textbf {\bibinfo {volume} {13}},\
  \bibinfo {pages} {376} (\bibinfo {year} {2017})}\BibitemShut {NoStop}%
\bibitem [{\citenamefont {Rein}\ and\ \citenamefont {Speck}(2016)}]{Rein2016}%
  \BibitemOpen
  \bibfield  {author} {\bibinfo {author} {\bibfnamefont {M.}~\bibnamefont
  {Rein}}\ and\ \bibinfo {author} {\bibfnamefont {T.}~\bibnamefont {Speck}},\
  }\href {https://doi.org/10.1140/epje/i2016-16084-7} {\bibfield  {journal}
  {\bibinfo  {journal} {Eur. Phys. J. E}\ }\textbf {\bibinfo {volume} {39}},\
  \bibinfo {pages} {84} (\bibinfo {year} {2016})}\BibitemShut {NoStop}%
\bibitem [{\citenamefont {Zakine}\ \emph {et~al.}(2020)\citenamefont {Zakine},
  \citenamefont {Zhao}, \citenamefont {Kne\ifmmode \check{z}\else
  \v{z}\fi{}evi\ifmmode~\acute{c}\else \'{c}\fi{}}, \citenamefont {Daerr},
  \citenamefont {Kafri}, \citenamefont {Tailleur},\ and\ \citenamefont {{Van
  Wijland}}}]{Zakine2020}%
  \BibitemOpen
  \bibfield  {author} {\bibinfo {author} {\bibfnamefont {R.}~\bibnamefont
  {Zakine}}, \bibinfo {author} {\bibfnamefont {Y.}~\bibnamefont {Zhao}},
  \bibinfo {author} {\bibfnamefont {M.}~\bibnamefont {Kne\ifmmode
  \check{z}\else \v{z}\fi{}evi\ifmmode~\acute{c}\else \'{c}\fi{}}}, \bibinfo
  {author} {\bibfnamefont {A.}~\bibnamefont {Daerr}}, \bibinfo {author}
  {\bibfnamefont {Y.}~\bibnamefont {Kafri}}, \bibinfo {author} {\bibfnamefont
  {J.}~\bibnamefont {Tailleur}},\ and\ \bibinfo {author} {\bibfnamefont
  {F.}~\bibnamefont {{Van Wijland}}},\ }\href@noop {} {\bibfield  {journal}
  {\bibinfo  {journal} {Phys. Rev. Lett.}\ }\textbf {\bibinfo {volume} {124}},\
  \bibinfo {pages} {248003}
  (\bibinfo {year} {2020})}\BibitemShut {NoStop}%
\bibitem [{\citenamefont {Marconi}\ \emph {et~al.}(2016)\citenamefont
  {Marconi}, \citenamefont {Maggi},\ and\ \citenamefont
  {Melchionna}}]{marconi_pressure_2016}%
  \BibitemOpen
  \bibfield  {author} {\bibinfo {author} {\bibfnamefont {U.~M.~B.}\
  \bibnamefont {Marconi}}, \bibinfo {author} {\bibfnamefont {C.}~\bibnamefont
  {Maggi}},\ and\ \bibinfo {author} {\bibfnamefont {S.}~\bibnamefont
  {Melchionna}},\ }\href {https://doi.org/10.1039/c6sm00667a} {\bibfield
  {journal} {\bibinfo  {journal} {Soft Matter}\ }\textbf {\bibinfo {volume}
  {12}},\ \bibinfo {pages} {5727} (\bibinfo {year} {2016})}\BibitemShut
  {NoStop}%
\bibitem [{\citenamefont {Speck}(2016)}]{Speck2016}%
  \BibitemOpen
  \bibfield  {author} {\bibinfo {author} {\bibfnamefont {T.}~\bibnamefont
  {Speck}},\ }\href {https://doi.org/10.1209/0295-5075/114/30006} {\bibfield
  {journal} {\bibinfo  {journal} {Europhys. Lett.}\ }\textbf {\bibinfo {volume}
  {114}},\ \bibinfo {pages} {30006} (\bibinfo {year} {2016})}\BibitemShut
  {NoStop}%
\bibitem [{\citenamefont {Marconi}\ and\ \citenamefont
  {Maggi}(2015)}]{MariniBettoloMarconi2015}%
  \BibitemOpen
  \bibfield  {author} {\bibinfo {author} {\bibfnamefont {U.~M.~B.}\
  \bibnamefont {Marconi}}\ and\ \bibinfo {author} {\bibfnamefont
  {C.}~\bibnamefont {Maggi}},\ }\href {https://doi.org/10.1039/c5sm01718a}
  {\bibfield  {journal} {\bibinfo  {journal} {Soft Matter}\ }\textbf {\bibinfo
  {volume} {11}},\ \bibinfo {pages} {8768} (\bibinfo {year}
  {2015})}\BibitemShut {NoStop}%
\bibitem [{\citenamefont {Wittmann}\ \emph {et~al.}(2019)\citenamefont
  {Wittmann}, \citenamefont {Smallenburg},\ and\ \citenamefont
  {Brader}}]{Wittmann2019}%
  \BibitemOpen
  \bibfield  {author} {\bibinfo {author} {\bibfnamefont {R.}~\bibnamefont
  {Wittmann}}, \bibinfo {author} {\bibfnamefont {F.}~\bibnamefont
  {Smallenburg}},\ and\ \bibinfo {author} {\bibfnamefont {J.~M.}\ \bibnamefont
  {Brader}},\ }\href {https://doi.org/10.1063/1.5086390} {\bibfield  {journal}
  {\bibinfo  {journal} {J. Chem. Phys.}\ }\textbf {\bibinfo {volume} {150}},\
  \bibinfo {pages} {174908} (\bibinfo {year} {2019})}\BibitemShut {NoStop}%
\bibitem [{\citenamefont {Marconi}\ \emph
  {et~al.}(2017{\natexlab{b}})\citenamefont {Marconi}, \citenamefont
  {Puglisi},\ and\ \citenamefont {Maggi}}]{Marconi2017}%
  \BibitemOpen
  \bibfield  {author} {\bibinfo {author} {\bibfnamefont {U.~M.~B.}\
  \bibnamefont {Marconi}}, \bibinfo {author} {\bibfnamefont {A.}~\bibnamefont
  {Puglisi}},\ and\ \bibinfo {author} {\bibfnamefont {C.}~\bibnamefont
  {Maggi}},\ }\href {https://doi.org/10.1038/srep46496} {\bibfield  {journal}
  {\bibinfo  {journal} {Sci. Rep.}\ }\textbf {\bibinfo {volume} {7}},\ \bibinfo
  {pages} {46496} (\bibinfo {year} {2017}{\natexlab{b}})}\BibitemShut {NoStop}%
\bibitem [{\citenamefont {Rodenburg}\ \emph {et~al.}(2017)\citenamefont
  {Rodenburg}, \citenamefont {Dijkstra},\ and\ \citenamefont {{Van
  Roij}}}]{Rodenburg2017}%
  \BibitemOpen
  \bibfield  {author} {\bibinfo {author} {\bibfnamefont {J.}~\bibnamefont
  {Rodenburg}}, \bibinfo {author} {\bibfnamefont {M.}~\bibnamefont
  {Dijkstra}},\ and\ \bibinfo {author} {\bibfnamefont {R.}~\bibnamefont {{Van
  Roij}}},\ }\href {https://doi.org/10.1039/c7sm01432e} {\bibfield  {journal}
  {\bibinfo  {journal} {Soft Matter}\ }\textbf {\bibinfo {volume} {13}},\
  \bibinfo {pages} {8957} (\bibinfo {year} {2017})}\BibitemShut {NoStop}%
\bibitem [{\citenamefont {Chakraborti}\ \emph {et~al.}(2016)\citenamefont
  {Chakraborti}, \citenamefont {Mishra},\ and\ \citenamefont
  {Pradhan}}]{Chakraborti2016}%
  \BibitemOpen
  \bibfield  {author} {\bibinfo {author} {\bibfnamefont {S.}~\bibnamefont
  {Chakraborti}}, \bibinfo {author} {\bibfnamefont {S.}~\bibnamefont
  {Mishra}},\ and\ \bibinfo {author} {\bibfnamefont {P.}~\bibnamefont
  {Pradhan}},\ }\href {https://doi.org/10.1103/PhysRevE.93.052606} {\bibfield
  {journal} {\bibinfo  {journal} {Phys. Rev. E}\ }\textbf {\bibinfo {volume}
  {93}},\ \bibinfo {pages} {052606} (\bibinfo {year} {2016})}\BibitemShut
  {NoStop}%
\bibitem [{\citenamefont {Takatori}\ and\ \citenamefont
  {Brady}(2015{\natexlab{a}})}]{Takatori2015}%
  \BibitemOpen
  \bibfield  {author} {\bibinfo {author} {\bibfnamefont {S.~C.}\ \bibnamefont
  {Takatori}}\ and\ \bibinfo {author} {\bibfnamefont {J.~F.}\ \bibnamefont
  {Brady}},\ }\href {https://doi.org/10.1103/PhysRevE.91.032117} {\bibfield
  {journal} {\bibinfo  {journal} {Phys. Rev. E}\ }\textbf {\bibinfo {volume}
  {91}},\ \bibinfo {pages} {32117} (\bibinfo {year}
  {2015}{\natexlab{a}})}\BibitemShut {NoStop}%
\bibitem [{\citenamefont {Paliwal}\ \emph {et~al.}(2018)\citenamefont
  {Paliwal}, \citenamefont {Rodenburg}, \citenamefont {{Van Roij}},\ and\
  \citenamefont {Dijkstra}}]{Paliwal2018}%
  \BibitemOpen
  \bibfield  {author} {\bibinfo {author} {\bibfnamefont {S.}~\bibnamefont
  {Paliwal}}, \bibinfo {author} {\bibfnamefont {J.}~\bibnamefont {Rodenburg}},
  \bibinfo {author} {\bibfnamefont {R.}~\bibnamefont {{Van Roij}}},\ and\
  \bibinfo {author} {\bibfnamefont {M.}~\bibnamefont {Dijkstra}},\ }\href
  {https://doi.org/10.1088/1367-2630/aa9b4d} {\bibfield  {journal} {\bibinfo
  {journal} {New J. Phys.}\ }\textbf {\bibinfo {volume} {20}},\ \bibinfo
  {pages} {015003} (\bibinfo {year} {2018})}\BibitemShut {NoStop}%
\bibitem [{\citenamefont {Wittkowski}\ \emph {et~al.}(2014)\citenamefont
  {Wittkowski}, \citenamefont {Tiribocchi}, \citenamefont {Stenhammar},
  \citenamefont {Allen}, \citenamefont {Marenduzzo},\ and\ \citenamefont
  {Cates}}]{Wittkowski2014}%
  \BibitemOpen
  \bibfield  {author} {\bibinfo {author} {\bibfnamefont {R.}~\bibnamefont
  {Wittkowski}}, \bibinfo {author} {\bibfnamefont {A.}~\bibnamefont
  {Tiribocchi}}, \bibinfo {author} {\bibfnamefont {J.}~\bibnamefont
  {Stenhammar}}, \bibinfo {author} {\bibfnamefont {R.~J.}\ \bibnamefont
  {Allen}}, \bibinfo {author} {\bibfnamefont {D.}~\bibnamefont {Marenduzzo}},\
  and\ \bibinfo {author} {\bibfnamefont {M.~E.}\ \bibnamefont {Cates}},\ }\href
  {https://doi.org/10.1038/ncomms5351} {\bibfield  {journal} {\bibinfo
  {journal} {Nat. Commun.}\ }\textbf {\bibinfo {volume} {5}},\ \bibinfo {pages}
  {4351} (\bibinfo {year} {2014})}\BibitemShut {NoStop}%
\bibitem [{\citenamefont {Takatori}\ and\ \citenamefont
  {Brady}(2015{\natexlab{b}})}]{Takatori2015a}%
  \BibitemOpen
  \bibfield  {author} {\bibinfo {author} {\bibfnamefont {S.~C.}\ \bibnamefont
  {Takatori}}\ and\ \bibinfo {author} {\bibfnamefont {J.~F.}\ \bibnamefont
  {Brady}},\ }\href {https://doi.org/10.1039/c5sm01792k} {\bibfield  {journal}
  {\bibinfo  {journal} {Soft Matter}\ }\textbf {\bibinfo {volume} {11}},\
  \bibinfo {pages} {7920} (\bibinfo {year} {2015}{\natexlab{b}})}\BibitemShut
  {NoStop}%
\bibitem [{\citenamefont {Partridge}\ and\ \citenamefont
  {Lee}(2019)}]{Partridge2019}%
  \BibitemOpen
  \bibfield  {author} {\bibinfo {author} {\bibfnamefont {B.}~\bibnamefont
  {Partridge}}\ and\ \bibinfo {author} {\bibfnamefont {C.~F.}\ \bibnamefont
  {Lee}},\ }\href {http://arxiv.org/abs/1810.06112} {\bibfield  {journal}
  {\bibinfo  {journal} {Phys. Rev. Lett.}\ }\textbf {\bibinfo {volume} {123}}, {\bibinfo {pages} {068002}} (\bibinfo {year} {2019})}\BibitemShut {NoStop}%
\bibitem [{\citenamefont {Solon}\ \emph {et~al.}(2018)\citenamefont {Solon},
  \citenamefont {Stenhammar}, \citenamefont {Cates}, \citenamefont {Kafri},\
  and\ \citenamefont {Tailleur}}]{Solon2018}%
  \BibitemOpen
  \bibfield  {author} {\bibinfo {author} {\bibfnamefont {A.~P.}\ \bibnamefont
  {Solon}}, \bibinfo {author} {\bibfnamefont {J.}~\bibnamefont {Stenhammar}},
  \bibinfo {author} {\bibfnamefont {M.~E.}\ \bibnamefont {Cates}}, \bibinfo
  {author} {\bibfnamefont {Y.}~\bibnamefont {Kafri}},\ and\ \bibinfo {author}
  {\bibfnamefont {J.}~\bibnamefont {Tailleur}},\ }\href
  {https://doi.org/10.1088/1367-2630/aaccdd} {\bibfield  {journal} {\bibinfo
  {journal} {New J. Phys.}\ }\textbf {\bibinfo {volume} {20}},\ \bibinfo
  {pages} {075001} (\bibinfo {year} {2018})}\BibitemShut {NoStop}%
\bibitem [{\citenamefont {Hermann}\ \emph
  {et~al.}(2019{\natexlab{a}})\citenamefont {Hermann}, \citenamefont
  {Krinninger}, \citenamefont {{de las Heras}},\ and\ \citenamefont
  {Schmidt}}]{Hermann2019}%
  \BibitemOpen
  \bibfield  {author} {\bibinfo {author} {\bibfnamefont {S.}~\bibnamefont
  {Hermann}}, \bibinfo {author} {\bibfnamefont {P.}~\bibnamefont {Krinninger}},
  \bibinfo {author} {\bibfnamefont {D.}~\bibnamefont {de las Heras}},\ and\
  \bibinfo {author} {\bibfnamefont {M.}~\bibnamefont {Schmidt}},\ }\href
  {https://doi.org/10.1103/PhysRevE.100.052604} {\bibfield  {journal} {\bibinfo
   {journal} {Phys. Rev. E}\ }\textbf {\bibinfo {volume} {100}},\ \bibinfo
  {pages} {52604} (\bibinfo {year} {2019}{\natexlab{a}})}\BibitemShut {NoStop}%
\bibitem [{\citenamefont {Levis}\ \emph {et~al.}(2017)\citenamefont {Levis},
  \citenamefont {Codina},\ and\ \citenamefont {Pagonabarraga}}]{Levis2017}%
  \BibitemOpen
  \bibfield  {author} {\bibinfo {author} {\bibfnamefont {D.}~\bibnamefont
  {Levis}}, \bibinfo {author} {\bibfnamefont {J.}~\bibnamefont {Codina}},\ and\
  \bibinfo {author} {\bibfnamefont {I.}~\bibnamefont {Pagonabarraga}},\ }\href
  {https://doi.org/10.1039/c7sm01504f} {\bibfield  {journal} {\bibinfo
  {journal} {Soft Matter}\ }\textbf {\bibinfo {volume} {13}},\ \bibinfo {pages}
  {8113} (\bibinfo {year} {2017})}\BibitemShut {NoStop}%
\bibitem [{\citenamefont {Hermann}\ \emph
  {et~al.}(2019{\natexlab{b}})\citenamefont {Hermann}, \citenamefont {{de las
  Heras}},\ and\ \citenamefont {Schmidt}}]{Hermann2019a}%
  \BibitemOpen
  \bibfield  {author} {\bibinfo {author} {\bibfnamefont {S.}~\bibnamefont
  {Hermann}}, \bibinfo {author} {\bibfnamefont {D.}~\bibnamefont {{de las
  Heras}}},\ and\ \bibinfo {author} {\bibfnamefont {M.}~\bibnamefont
  {Schmidt}},\ }\href {https://doi.org/10.1103/PhysRevLett.123.268002}
  {\bibfield  {journal} {\bibinfo  {journal} {Phys. Rev. Lett.}\ }\textbf
  {\bibinfo {volume} {123}},\ \bibinfo {pages} {268002} (\bibinfo {year}
  {2019}{\natexlab{b}})}\BibitemShut {NoStop}%
\bibitem [{\citenamefont {Tjhung}\ \emph {et~al.}(2018)\citenamefont {Tjhung},
  \citenamefont {Nardini},\ and\ \citenamefont {Cates}}]{Tjhung2018}%
  \BibitemOpen
  \bibfield  {author} {\bibinfo {author} {\bibfnamefont {E.}~\bibnamefont
  {Tjhung}}, \bibinfo {author} {\bibfnamefont {C.}~\bibnamefont {Nardini}},\
  and\ \bibinfo {author} {\bibfnamefont {M.~E.}\ \bibnamefont {Cates}},\ }\href
  {https://doi.org/10.1103/PhysRevX.8.031080} {\bibfield  {journal} {\bibinfo
  {journal} {Phys. Rev. X}\ }\textbf {\bibinfo {volume} {8}},\ \bibinfo {pages}
  {031080} (\bibinfo {year} {2018})}\BibitemShut {NoStop}%
\bibitem [{\citenamefont {de~Gennes}(1979)}]{Gennes1979}%
  \BibitemOpen
  \bibfield  {author} {\bibinfo {author} {\bibfnamefont {P.-G.}\ \bibnamefont
  {de~Gennes}},\ }\href {https://books.google.com/books?id=ApzfJ2LYwGUC} {\emph
  {\bibinfo {title} {Scaling Concepts in Polymer Physics}}}\ (\bibinfo
  {publisher} {Cornell University Press},\ \bibinfo {year} {1979})\BibitemShut
  {NoStop}%
\bibitem [{\citenamefont {Rubinstein}\ and\ \citenamefont
  {Colby}(2003)}]{Rubinstein2003}%
  \BibitemOpen
  \bibfield  {author} {\bibinfo {author} {\bibfnamefont {M.}~\bibnamefont
  {Rubinstein}}\ and\ \bibinfo {author} {\bibfnamefont {R.}~\bibnamefont
  {Colby}},\ }\href@noop {} {\emph {\bibinfo {title} {Polymer Physics}}}\
  (\bibinfo  {publisher} {OUP Oxford},\ \bibinfo {year} {2003})\BibitemShut
  {NoStop}%
\bibitem [{\citenamefont {Wang}(2017)}]{Wang2017}%
  \BibitemOpen
  \bibfield  {author} {\bibinfo {author} {\bibfnamefont {Z.-G.}\ \bibnamefont
  {Wang}},\ }\href {https://doi.org/10.1021/acs.macromol.7b01518} {\bibfield
  {journal} {\bibinfo  {journal} {Macromolecules}\ }\textbf {\bibinfo {volume}
  {50}},\ \bibinfo {pages} {9073} (\bibinfo {year} {2017})}\BibitemShut
  {NoStop}%
\bibitem [{\citenamefont {Digregorio}\ \emph {et~al.}(2018)\citenamefont
  {Digregorio}, \citenamefont {Levis}, \citenamefont {Suma}, \citenamefont
  {Cugliandolo}, \citenamefont {Gonnella},\ and\ \citenamefont
  {Pagonabarraga}}]{Digregorio2018}%
  \BibitemOpen
  \bibfield  {author} {\bibinfo {author} {\bibfnamefont {P.}~\bibnamefont
  {Digregorio}}, \bibinfo {author} {\bibfnamefont {D.}~\bibnamefont {Levis}},
  \bibinfo {author} {\bibfnamefont {A.}~\bibnamefont {Suma}}, \bibinfo {author}
  {\bibfnamefont {L.~F.}\ \bibnamefont {Cugliandolo}}, \bibinfo {author}
  {\bibfnamefont {G.}~\bibnamefont {Gonnella}},\ and\ \bibinfo {author}
  {\bibfnamefont {I.}~\bibnamefont {Pagonabarraga}},\ }\href
  {https://doi.org/10.1103/PhysRevLett.121.098003} {\bibfield  {journal}
  {\bibinfo  {journal} {Phys. Rev. Lett.}\ }\textbf {\bibinfo {volume} {121}},\
  \bibinfo {pages} {098003} (\bibinfo {year} {2018})}\BibitemShut {NoStop}%
\bibitem [{\citenamefont {Klamser}\ \emph {et~al.}(2018)\citenamefont
  {Klamser}, \citenamefont {Kapfer},\ and\ \citenamefont
  {Krauth}}]{Klamser2018}%
  \BibitemOpen
  \bibfield  {author} {\bibinfo {author} {\bibfnamefont {J.~U.}\ \bibnamefont
  {Klamser}}, \bibinfo {author} {\bibfnamefont {S.~C.}\ \bibnamefont
  {Kapfer}},\ and\ \bibinfo {author} {\bibfnamefont {W.}~\bibnamefont
  {Krauth}},\ }\href {https://doi.org/10.1038/s41467-018-07491-5} {\bibfield
  {journal} {\bibinfo  {journal} {Nat. Commun.}\ }\textbf {\bibinfo {volume}
  {9}},\ \bibinfo {pages} {5045} (\bibinfo {year} {2018})}\BibitemShut
  {NoStop}%
\bibitem [{\citenamefont {Paliwal}\ and\ \citenamefont
  {Dijkstra}(2020)}]{Paliwal2020}%
  \BibitemOpen
  \bibfield  {author} {\bibinfo {author} {\bibfnamefont {S.}~\bibnamefont
  {Paliwal}}\ and\ \bibinfo {author} {\bibfnamefont {M.}~\bibnamefont
  {Dijkstra}},\ }\href {https://doi.org/10.1103/PhysRevResearch.2.012013}
  {\bibfield  {journal} {\bibinfo  {journal} {Phys. Rev. Res.}\ }\textbf
  {\bibinfo {volume} {2}},\ \bibinfo {pages} {012013} (\bibinfo {year}
  {2020})}\BibitemShut {NoStop}%
\bibitem [{\citenamefont {Weeks}\ \emph {et~al.}(1971)\citenamefont {Weeks},
  \citenamefont {Chandler},\ and\ \citenamefont {Andersen}}]{Weeks2011}%
  \BibitemOpen
  \bibfield  {author} {\bibinfo {author} {\bibfnamefont {J.~D.}\ \bibnamefont
  {Weeks}}, \bibinfo {author} {\bibfnamefont {D.}~\bibnamefont {Chandler}},\
  and\ \bibinfo {author} {\bibfnamefont {H.~C.}\ \bibnamefont {Andersen}},\
  }\href {https://doi.org/10.1063/1.1674820} {\bibfield  {journal} {\bibinfo
  {journal} {J. Chem. Phys.}\ }\textbf {\bibinfo {volume} {54}},\ \bibinfo
  {pages} {5237} (\bibinfo {year} {1971})}\BibitemShut {NoStop}%
\bibitem [{\citenamefont {Anderson}\ \emph {et~al.}(2020)\citenamefont
  {Anderson}, \citenamefont {Glaser},\ and\ \citenamefont
  {Glotzer}}]{Anderson2020}%
  \BibitemOpen
  \bibfield  {author} {\bibinfo {author} {\bibfnamefont {J.~A.}\ \bibnamefont
  {Anderson}}, \bibinfo {author} {\bibfnamefont {J.}~\bibnamefont {Glaser}},\
  and\ \bibinfo {author} {\bibfnamefont {S.~C.}\ \bibnamefont {Glotzer}},\
  }\href {https://doi.org/10.1016/j.commatsci.2019.109363} {\bibfield
  {journal} {\bibinfo  {journal} {Comput. Mater. Sci.}\ }\textbf {\bibinfo
  {volume} {173}},\ \bibinfo {pages} {109363} (\bibinfo {year}
  {2020})}\BibitemShut {NoStop}%
\bibitem [{\citenamefont {Bickmann}\ and\ \citenamefont
  {Wittkowski}(2020{\natexlab{a}})}]{Bickmann2020a}%
  \BibitemOpen
  \bibfield  {author} {\bibinfo {author} {\bibfnamefont {J.}~\bibnamefont
  {Bickmann}}\ and\ \bibinfo {author} {\bibfnamefont {R.}~\bibnamefont
  {Wittkowski}},\ }\href@noop {} {\bibfield  {journal} {\bibinfo  {journal}
  {Phys. Rev. Res.}\ }\textbf {\bibinfo {volume} {2}}, {\bibinfo {pages} {033241}} (\bibinfo {year}
  {2020}{\natexlab{a}})}\BibitemShut {NoStop}%
\bibitem [{\citenamefont {Bickmann}\ and\ \citenamefont
  {Wittkowski}(2020{\natexlab{b}})}]{Bickmann2020}%
  \BibitemOpen
  \bibfield  {author} {\bibinfo {author} {\bibfnamefont {J.}~\bibnamefont
  {Bickmann}}\ and\ \bibinfo {author} {\bibfnamefont {R.}~\bibnamefont
  {Wittkowski}},\ }\href {https://doi.org/10.1088/1361-648X/ab5e0e} {\bibfield
  {journal} {\bibinfo  {journal} {J. Phys. Condens. Matter}\ }\textbf {\bibinfo
  {volume} {32}},\ \bibinfo {pages} {214001} (\bibinfo {year}
  {2020}{\natexlab{b}})}\BibitemShut {NoStop}%
\bibitem [{\citenamefont {Jeggle}\ \emph {et~al.}(2020)\citenamefont {Jeggle},
  \citenamefont {Stenhammar},\ and\ \citenamefont {Wittkowski}}]{Jeggle2020}%
  \BibitemOpen
  \bibfield  {author} {\bibinfo {author} {\bibfnamefont {J.}~\bibnamefont
  {Jeggle}}, \bibinfo {author} {\bibfnamefont {J.}~\bibnamefont {Stenhammar}},\
  and\ \bibinfo {author} {\bibfnamefont {R.}~\bibnamefont {Wittkowski}},\
  }\href {https://doi.org/10.1063/1.5140725} {\bibfield  {journal} {\bibinfo
  {journal} {J. Chem. Phys.}\ }\textbf {\bibinfo {volume} {152}},\ \bibinfo
  {pages} {194903} (\bibinfo {year} {2020})}\BibitemShut {NoStop}%
\bibitem [{\citenamefont {Squires}\ and\ \citenamefont
  {Brady}(2005)}]{Squires2005}%
  \BibitemOpen
  \bibfield  {author} {\bibinfo {author} {\bibfnamefont {T.~M.}\ \bibnamefont
  {Squires}}\ and\ \bibinfo {author} {\bibfnamefont {J.~F.}\ \bibnamefont
  {Brady}},\ }\href@noop {} {\bibfield  {journal} {\bibinfo  {journal} {Phys.
  Fluids}\ }\textbf {\bibinfo {volume} {17}},\ \bibinfo {pages} {073101}
  (\bibinfo {year} {2005})}\BibitemShut {NoStop}%
\bibitem [{Note2()}]{Note2}%
  \BibitemOpen
  \bibinfo {note} {See [URL], which includes Refs.~\cite {Takatori2015,
  Solon2018, Santos2017}, for supporting data for the scaling analysis
  presented in the main text and the precise functional forms of the
  equations-of-state.}\BibitemShut {Stop}%
\bibitem [{\citenamefont {Santos}\ \emph {et~al.}(2017)\citenamefont {Santos},
  \citenamefont {Yuste}, \citenamefont {L\'opez~de Haro},\ and\ \citenamefont
  {Ogarko}}]{Santos2017}%
  \BibitemOpen
  \bibfield  {author} {\bibinfo {author} {\bibfnamefont {A.}~\bibnamefont
  {Santos}}, \bibinfo {author} {\bibfnamefont {S.~B.}\ \bibnamefont {Yuste}},
  \bibinfo {author} {\bibfnamefont {M.}~\bibnamefont {L\'opez~de Haro}},\ and\
  \bibinfo {author} {\bibfnamefont {V.}~\bibnamefont {Ogarko}},\ }\href
  {https://doi.org/10.1103/PhysRevE.96.062603} {\bibfield  {journal} {\bibinfo
  {journal} {Phys. Rev. E}\ }\textbf {\bibinfo {volume} {96}},\ \bibinfo
  {pages} {062603} (\bibinfo {year} {2017})}\BibitemShut {NoStop}%
\end{thebibliography}
%
\end{document}


\title{Supplemental Material: \\ Dynamic Overlap Concentration Scale of Active Colloids}

\author{Stewart A. Mallory}
\email{smallory@caltech.edu}
\affiliation{Division of Chemistry and Chemical Engineering, California Institute of Technology, Pasadena, California, 91125, USA}

\author{Ahmad K. Omar}
\email{aomar@berkeley.edu}
\affiliation{Department of Chemistry, University of California, Berkeley, California, 94720, USA }

\author{John F. Brady}
\email{jfbrady@caltech.edu}
\affiliation{Division of Chemistry and Chemical Engineering, California Institute of Technology, Pasadena, California, 91125, USA}

\maketitle

\section{Identifying the dynamic overlap concentration}

\begin{figure*}[h!]
	\centering
	\includegraphics[width=1.0\textwidth]{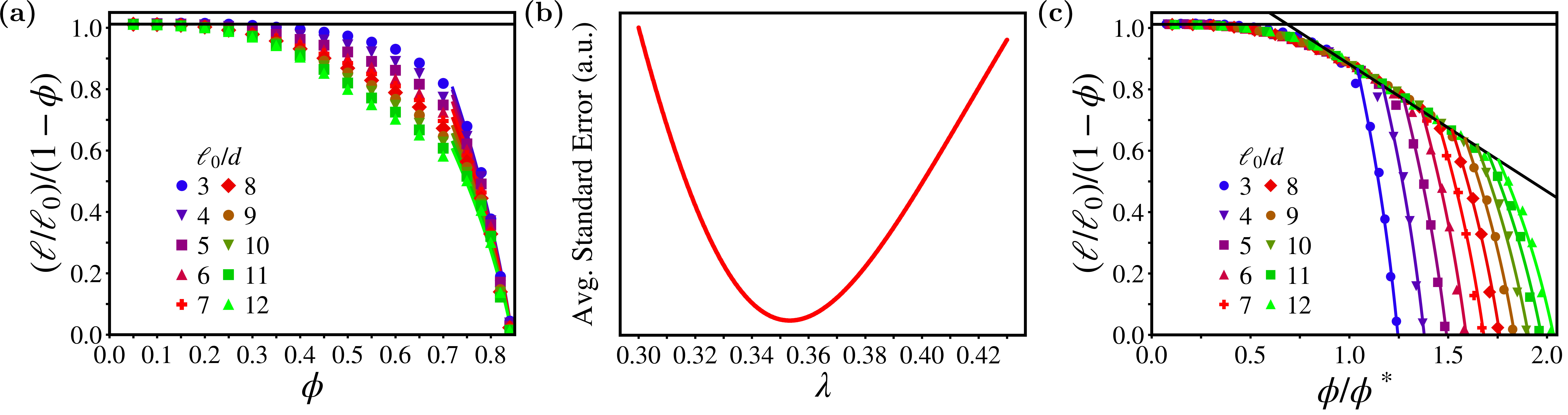}
	\caption{\protect\small{{(a) Normalized run-length rescaled by the first-order correction due to particle interactions. The dilute regime is demarcated by the black horizontal line and the family of curves in the active jamming regime have the functional form described in the text. (b) Minimization of the average standard error in the fitting parameters to determine the optimal value of $\lambda$ for the overlap concentration $\phi^*=(d/\ell_0)^\lambda$. (c) Collapse of the dilute and semidilute regions for the effective run-length using the determined dynamic overlap concentration $\phi^*$. The colored lines represent our theoretical equations-of-state in the active jamming region, while the black lines characterize the dilute and semidilute regimes. }}}
	\label{SM_figure1}
\end{figure*}

In this section, we provide the technical details for identifying the dynamic overlap concentration $\phi^*$ for the polydisperse suspension of active Brownian particles (ABPs) considered in this study. 
This procedure collapses the normalized run-length data shown in Fig.~\ref{SM_figure1}(a) onto the set of principal curves shown in Fig.~\ref{SM_figure1}(c), corresponding to Figs. 2(b) and 2(c), respectively, in the main text.

As discussed in the main text, the first-order correction to the effective run-length due to interparticle collisions is given by $\ell/\ell_0=\Pi^s/\Pi^s_0=(1-\phi)$.
This expression captures the behavior of the effective run-length (which is directly proportional to the swim pressure) at low concentrations as shown in Fig.~\ref{SM_figure1}(a) (black horizontal line), while in the active jamming regime, we find $\ell$ is well-described by, $\ell/\ell_0=\Pi^s/\Pi^s_0=\mathcal{D}(1-e^{-\mathcal{E}/\ell_0})(1-\phi/\phi_m)$, where $\mathcal{D},\mathcal{E}>0$ are constant fitting parameters.
As shown by the family of curves in Fig.~\ref{SM_figure1}(c), values of $\mathcal{D}=0.55$ and $\mathcal{E}=16.5$ accurately capture the $\ell_0$ dependence of the normalized run-length in the active jamming regime.

The semidilute regime for each curve exhibits a linear decay with increasing concentration and the width of this region gradually broadens as the intrinsic run-length is increased.
The linear scaling in the semidilute regime suggest that all points in the region can be collapsed onto a single line by rescaling the $\phi$-axis by the overlap concentration $\phi^*=(d/\ell_0)^\lambda$.
To determine the optimal value of $\lambda$, we use an iterative procedure where all points in the intermediate region are fit using a simple linear regression for different values of $\lambda$ and identify the value of $\lambda$ that minimizes the average standard error of the two fitting parameters [see Fig.~\ref{SM_figure1}(b)]. 
The optimal value is found to be $\lambda \approx 0.353$ and the resulting linear function, which is shown in Fig. ~\ref{SM_figure1}(c), is $\mathcal{B} (\phi/\phi^*)+\mathcal{C}$ where $\mathcal{B}=-0.414$ and $\mathcal{C}=1.296$. 

\section{Construction of swim pressure}

\begin{figure*}[h!]
\centering
	\includegraphics[width=1.0\textwidth]{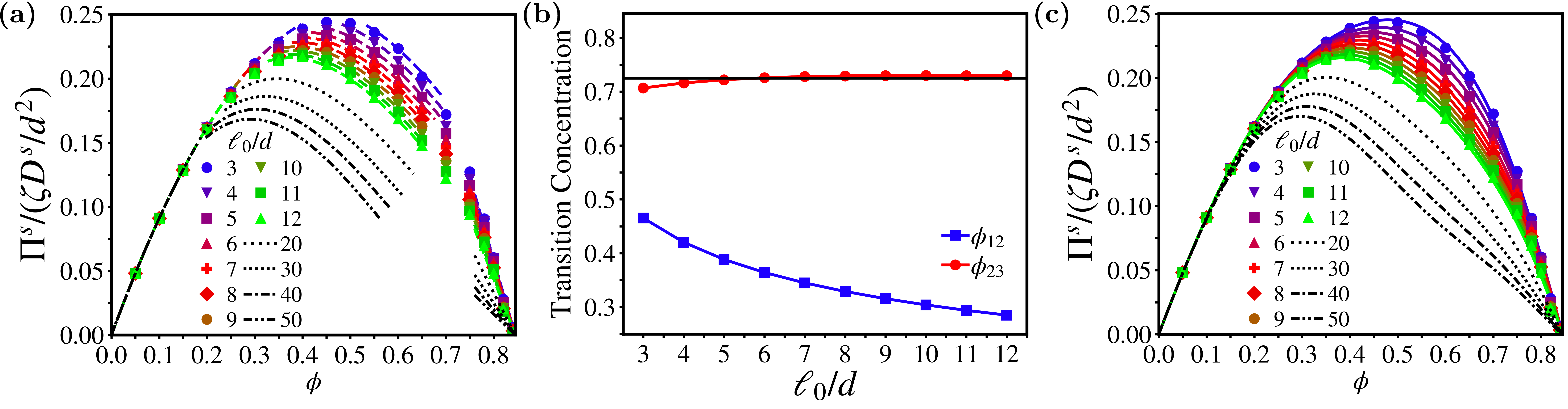}
	\caption{\protect\small{{(a) The piecewise expression of the swim pressure in both regions of homogeneity and inhomogeneity in comparison with simulation data. (b) Transition concentrations from the dilute to semidilute regime ($\phi_{12}$) and semidilute to active jamming regime ($\phi_{23}$). (c) Continuous expression for the swim pressure which is fit to the points obtained from the piecewise expression.}}}
	\label{SM_figure2}
\end{figure*}

The procedure outlined in Section I quantitatively characterizes the three regions of the swim pressure (dilute, semidilute and active jamming) and admits the construction of the piecewise expression:

\begin{equation}
\label{eq:swimpiece}
 \frac{\Pi^s}{\Pi^s_0}= \frac{\ell}{\ell_0} =
    \begin{cases} 
        (1-\phi) \mathcal{A}, & \phi \leq \phi_{12} - \delta \\
        (1-\phi) (\mathcal{B} (\phi/\phi^*)+\mathcal{C}), &  \phi_{12} + \delta <  \phi \leq \phi_{23} - \tilde{\delta} \\
      \mathcal{D}(1-e^{-\mathcal{E}/\ell_0})(1-\phi/\phi_m), & \phi_{23} +\delta < \phi \leq \phi_m 
   \end{cases}
\end{equation}

\noindent where $\mathcal{A}=1.012$ is a small empirical correction factor for the low-density region and $\mathcal{B}$, $\mathcal{C}$, $\mathcal{D}$, and $\mathcal{E}$ are the parameters determined in Section I. 
A comparison between this piecewise expression and swim pressure measurements from simulation are presented in Fig.~\ref{SM_figure2}(a) for $3 \leq \ell_0 \leq 12$. 
The transition concentrations between regimes were estimated by computing the point of intersection between the two functions describing consecutive regimes [see Fig.~\ref{SM_figure2}(b)].
The dilute to semidilute transition concentration is $\phi_{12} \approx 0.686 \phi^*$, while the transition concentration from the semidilute to the active jamming regime is nearly constant (independent of $\phi^*$) at $\phi_{23} \approx 0.72$.
The piecewise expression for the swim pressure describes points unambiguously belonging to the three concentration regimes. 
To exclude the ambiguous points between concentration regimes, concentrations within a factor $\delta = 0.1$ [see Eq.~\eqref{eq:swimpiece}] of a transition concentration are not described by Eq.~\eqref{eq:swimpiece} [see Fig.~\ref{SM_figure2}(a)]. 
Additionally, a region between the semidilute and active jamming region is excluded in the piecewise function to ensure that the swim pressure is a monotonically decreasing function in these two regions for all $\ell_0$ values considered.
The excluded region is determined such that the minimum value of the swim pressure in the semidilute regime is equal to $\Pi^s/\Pi^s_0=\ell/\ell_0=(\mathcal{D}(1-e^{-\mathcal{E}/\ell_0})(1-(\phi_{23})/\phi_m))+\Delta$ where $\Delta=0.2$ -- a condition that is straightforward to enforce numerically and results in a dynamic shift-factor $\tilde{\delta}$ [see Eq.~\eqref{eq:swimpiece}].
The results are insensitive to the choice of $\Delta$. 
The primary function of $\Delta$ is to ensure that for large values of $\ell_0$ the fitting procedure to produce a continuous function doesn't introducing spurious artifacts.

As swim pressure is continuous and differentiable for all values of $\phi$, we fit the above piecewise expression to the function:

\begin{equation}
\label{eq:swimfit}
    \frac{\Pi^s}{\Pi^s_0}=\frac{\ell}{\ell_0}=(1-\phi/\phi_m)(1+a_1\phi+a_2\phi^2+a_3\phi^3+a_4\phi^4+a_5\phi^5).
\end{equation}

\noindent This functional form captures all three regimes of the swim pressure described in Eq.~\eqref{eq:swimpiece} while ensuring a smooth transition between them.
The coefficients are determined by using a simple linear regression where the expression above is fit to the piecewise data for the swim pressure. 
The resulting continuous functions for the swim pressure are given in Fig.~\ref{SM_figure2}(c) and the associated fitting parameters for different values of $\ell_0$ are provided in Table I.
This is the procedure used to generate the swim pressure curves shown in Fig. 2(d) of the main text.

\section{Verifying overlap concentration}

\begin{figure*}[h!]
	\centering
	\includegraphics[width=1.0\textwidth]{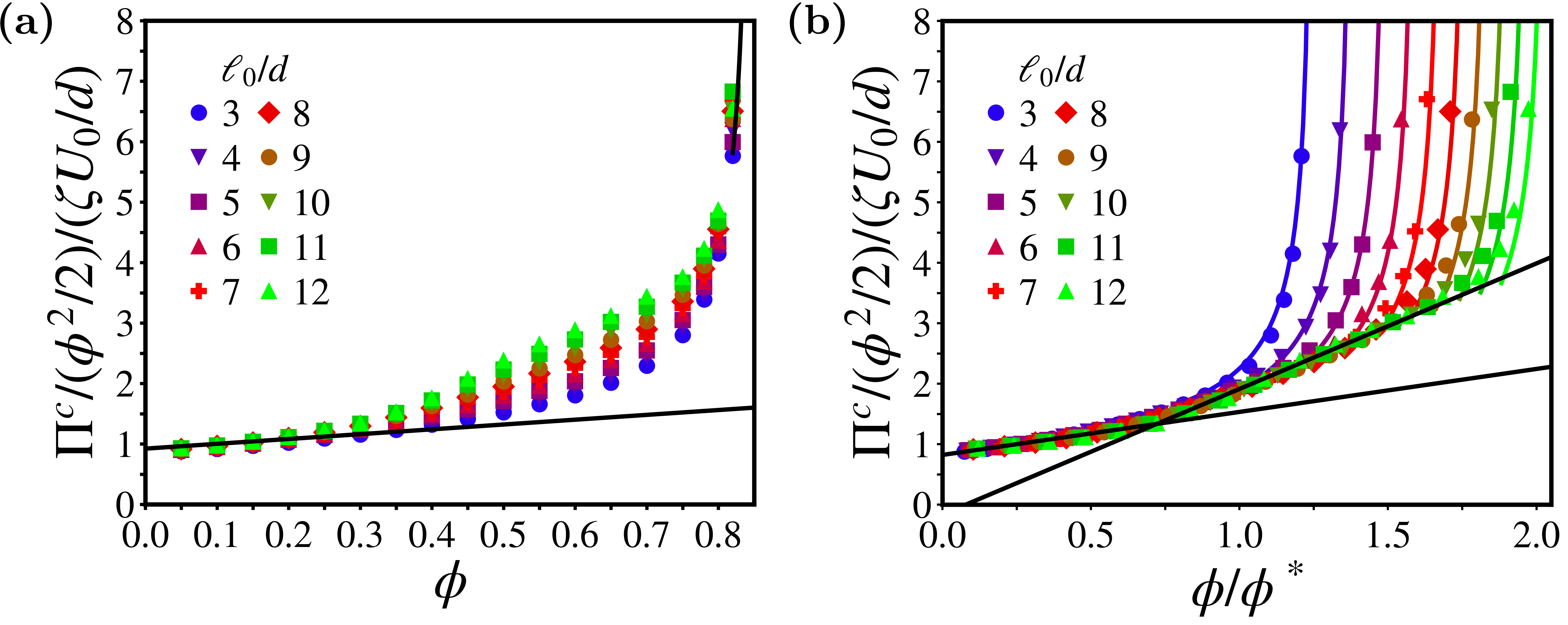}
	\caption{\protect\small{{(a) Dimensionless collisional pressure rescaled by the first order correction due to interparticle interactions $(\phi^2/2)$. The points in the dilute regime can be described by a line and the active jamming region near maximal-packing diverges as $\Pi^c \sim 1/(1-\phi/\phi_m)^{\frac{1}{2}}$. (b) Collapse of the dilute and semidilute regions for the collisonal pressure using the dynamic overlap concentration $\phi^*$. The colored lines represent our theoretical equations-of-state in the active jamming region, while the black lines characterize the dilute and semidilute regimes.}}}
	\label{SM_figure3}
\end{figure*}

The dynamic overlap concentration computed in Section I also collapses the collisional pressure onto a set of principal curves [see Fig.~\ref{SM_figure3}].
These are the details of the intermediate steps needed to collapse the curves in Fig. 2(f) into those of Fig. 2(g) of the main text.  

As shown in Fig.~\ref{SM_figure3}(a), the first order correction to the collisional pressure scales as $\Pi^c \sim (\phi^2/2)$. 
The rescaling of the collisional pressure $\Pi^c$ by $(\phi^2/2)$ collapses the dilute regime onto a line $\mathcal{F}\phi+\mathcal{G}$ where $\mathcal{F}=0.711$ and $\mathcal{G}=0.822$.
We find the asymptotic behavior of the collisional pressure in the active jamming regime is given by the functional form $\Pi^c \sim 1/(1-\phi/\phi_m)^{\frac{1}{2}}$.
This expression is similar to a recent EoS proposed by Santos et al.~\cite{Santos2017} for equilibrium polydisperse hard disks providing further evidence that at high concentrations system behavior becomes independent of $\ell_0$. 
Similar to the swim pressure, we observe a linear dependence of the rescaled collisional pressure on concentration in the semidilute regime (implying $\Pi^c \sim \phi^3$).
Using the value of $\lambda$ determined in Section I for the overlap concentration $\phi^*=(d/\ell_0)^\lambda$ it is possible to rescale the $\phi$-axis such that the intermediate region collapse onto the single line $\mathcal{H}(\phi/\phi^*)+\mathcal{I}$ where $\mathcal{H}=2.077$ and $\mathcal{I}=-0.163$ [see Fig.~\ref{SM_figure3}(b)].

\section{Construction of collisional pressure}

\begin{figure*}[h!]
	\centering
	\includegraphics[width=1.0\textwidth]{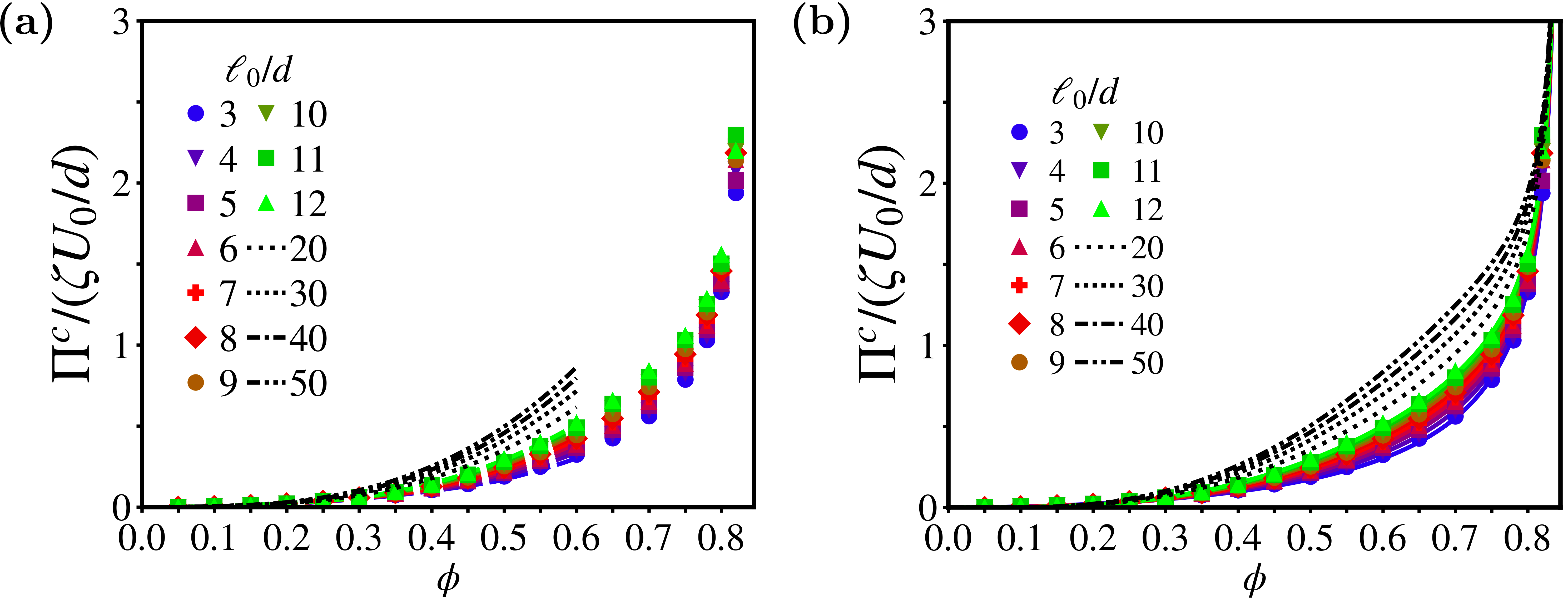}
	\caption{\protect\small{{(a) The piecewise expression of the collisional pressure in both regions of homogeneity and inhomogeneity in comparison with simulation data. (b) Continuous expression for the collisional pressure which is fit to the points obtained from the piecewise expression.}}}
	\label{SM_figure4}
\end{figure*}

The piecewise expression for the collisional pressure is:  

\begin{equation}
\label{eq:collpiece}
 \frac{\Pi^c}{(\phi^2/2)}=
    \begin{cases} 
        \mathcal{F}\phi+\mathcal{G}, & \phi \leq \phi_{12}^{\prime} \\
      \mathcal{H}(\phi/\phi^*)+\mathcal{I}, &  \phi_{12}^{\prime} <  \phi \leq \phi_{23}^{\prime}  \\
      1/(1-\phi/\phi_m)^{\frac{1}{2}}, &  \phi_{23}^{\prime} + \delta^{\prime} < \phi \leq \phi_m 
   \end{cases}
\end{equation}

\noindent where $\mathcal{F}$, $\mathcal{G}$, $\mathcal{H}$, and $\mathcal{I}$ are the parameters determined in Section III. 
A comparison between this piecewise expression and collisional pressure measurements from simulation is presented in Fig.~\ref{SM_figure4}(a) for $3 \leq \ell_0 \leq 12$. 
For the collisional pressure, the transition concentration between the dilute and semidilute regime $\phi_{12}^{\prime}$ was estimated by computing the point of intersection between the two functions describing these regions and determined to be $\phi_{12}^{\prime} \approx 0.722\phi^*$.
A transition concentration value of $\phi_{23}^{\prime}=0.6$ gives good agreement with simulation data when fitting to a continuous function.
We select a value of $\delta^{\prime}=0.24$ to capture values near maximal-packing [see Eq. \eqref{eq:collpiece}].

The continuous and differentiable functional form for the collisional pressure is given by:

\begin{equation}
\label{eq:collfit}
    \Pi^c=\frac{b_2\phi^2+b_3\phi^3+b_4\phi^4}{(1-\phi/\phi_m)^{\frac{1}{2}}}
\end{equation}

\noindent The coefficients are determined by using a simple linear regression where the expression is fit to the piecewise expression for the collisional pressure.
The resulting continuous functions for the collisional pressure are given in Fig.~\ref{SM_figure4}(b) and the associated fitting parameters for different values of $\ell_0$ are provided in Table II.
This is the procedure used to generate the collisional pressure curves shown in Fig. 2(h) of the main text.

\section{Details of Equations-of-State from Previous Works}

The equations-of-state adapted from the work of Solon et al.~\cite{Solon2018} have the following functional form: 

\begin{align*} 
\Pi^s & =\Pi^s_0(1+a_1\phi+a_2\phi^2)(1-\tanh \left(a_3(\phi-a_4) \right)), \\ 
\Pi^c & =b_0(e^{b_1\phi}-1)+b_2(e^{b_3 \phi}-1),
\end{align*}

\noindent where the fitting parameters to our swim pressure data are $a_1=-1.011$, $a_2=-0.1972$, $a_3=0.4476$, $a_4=0.2333$ and the fitting parameters to our collisonal pressure data are $b_0=0.0$, $b_1=88.84$, $b_2=0.03582$, $b_3=4.68$

The equations-of-state adapted from the work of Takatori et al.~\cite{Takatori2015} has the following functional form: 

\begin{align*} 
\Pi^s & =\Pi^s_0(1-\phi+a_2 \phi^2), \\ 
\Pi^c & =b_0/(1-\phi/\phi_m)^{b_1},
\end{align*}

\noindent where the fitting parameter to our swim pressure data is $a_2=-0.2537$ and the fitting parameters to our collisional pressure data are $b_0=0.4387$ and $b_1=0.5563$.

The fitting parameters for both equations-of-state were determined by fitting the appropriate function to the simulation data near the critical point for MIPS ($\ell_0=12$) - an analogous procedure to that implemented in each of these works.

\newpage

\begin{table}[h!]
\begin{ruledtabular}
\begin{tabular}{dddddd}
\ell_0 & a_1 & a_2 & a_3 & a_4 & a_5\\
\colrule
3.0  & 0.183432 & 2.845697 & -15.147852 & 29.725037 & -18.192493\\
4.0  & 0.183432 & 2.891293 & -14.835777 & 26.893139 & -15.248972\\
5.0  & 0.183432 & 2.702787 & -13.178302 & 21.966445 & -11.382876\\
6.0  & 0.183432 & 2.438215 & -11.141770 & 16.631632 & -7.536175\\
7.0  & 0.183432 & 2.499049 & -11.390424 & 16.064329 & -6.699915\\
8.0  & 0.183432 & 2.683782 & -12.597868 & 17.664652 & -7.321000\\
9.0  & 0.183432 & 2.907041 & -14.169293 & 20.196494 & -8.615149\\
10.0 & 0.183432 & 3.129855 & -15.821185 & 23.046509 & -10.179068\\
11.0 & 0.183432 & 3.334186 & -17.415974 & 25.905089 & -11.802772\\
12.0 & 0.183432 & 3.510066 & -18.875512 & 28.591585 & -13.361462\\
15.0 & 0.183432 & 3.848343 & -22.225547 & 35.012108 & -17.184935\\
20.0 & 0.183432 & 3.918075 & -24.893074 & 40.622870 & -20.650856\\
25.0 & 0.183432 & 3.690035 & -25.640354 & 42.716735 & -22.007869\\
30.0 & 0.183432 & 3.380314 & -25.740661 & 43.591748 & -22.603205\\
35.0 & 0.183432 & 3.071195 & -25.717216 & 44.237095 & -23.045548\\
40.0 & 0.183432 & 2.789162 & -25.760161 & 45.029119 & -23.578574\\
45.0 & 0.183432 & 2.534460 & -25.894524 & 46.027805 & -24.250557\\
50.0 & 0.183432 & 2.302974 & -26.109042 & 47.214032 & -25.055000\\
\end{tabular}
\end{ruledtabular}
	\caption{\protect\small{{Fitting parameters for the swim pressure [see Eq.~\eqref{eq:swimfit}] for various values of $\ell_0$. Note that $a_1$ is fixed at the appropriate value to reproduce the correct low density result $\Pi^s/\Pi^s_0=(1-\phi)$. The remaining coefficients are determined by using a simple linear regression where the expression is fit to the piecewise expression for swim pressure. }}}
\end{table}

\begin{table}[H]
\begin{ruledtabular}
\begin{tabular}{dddd}
\ell_0 & b_2 & b_3 & b_4 \\
\colrule
3.0  & 0.516712  & -0.326021 &  0.362842 \\
4.0  & 0.345719  & 0.449888  & -0.316453 \\
5.0  & 0.219327  & 1.061960  & -0.864242 \\
6.0  & 0.122881  & 1.558342  & -1.316998 \\
7.0  & 0.046163  & 1.974989  & -1.702971 \\
8.0  & -0.017101 & 2.334871  & -2.040573 \\
9.0  & -0.070816 & 2.652788  & -2.341859 \\
10.0 & -0.117500 & 2.938596  & -2.614970 \\
11.0 & -0.158844 & 3.199109  & -2.865605 \\
12.0 & -0.196018 & 3.439187  & -3.097882 \\
15.0 & -0.289852 & 4.068476  & -3.711784 \\
20.0 & -0.410507 & 4.920119  & -4.551497 \\
25.0 & -0.506271 & 5.622290  & -5.249047 \\
30.0 & -0.587207 & 6.228834  & -5.854106 \\
35.0 & -0.658161 & 6.767830  & -6.393142 \\
40.0 & -0.721849 & 7.255988  & -6.882142 \\
45.0 & -0.779960 & 7.704183  & -7.331619 \\
50.0 & -0.833623 & 8.119935  & -7.748900 \\
\end{tabular}
\end{ruledtabular}
	\caption{\protect\small{{Fitting parameters for the collisional pressure [see Eq.~\eqref{eq:collfit}] for various values of $\ell_0$. The coefficients are determined by using a simple linear regression where the expression is fit to the piecewise expression for collisional pressure.}}}
\end{table}